\documentclass{article}
\pdfoutput=1
\usepackage{jheppub}
\usepackage{bm}
\usepackage{braket}
\usepackage[table]{xcolor}
\usepackage{mathrsfs}
\usepackage[english]{babel}
\usepackage{graphicx,multirow}
\usepackage{caption}
\usepackage{subcaption}
\usepackage{amssymb}
\usepackage{makecell}
\usepackage{physics}
\usepackage{placeins}
\usepackage{hhline}
\usepackage{tcolorbox}
\usepackage{upgreek}
\usepackage[makeroom]{cancel}

\makeatletter
\gdef\@fpheader{}
\g@addto@macro\bfseries{\boldmath}
\makeatother

\def\mfa{{\mathfrak{a}}}

\def\exd{{\hbox{d}}}

\def\bea{\begin{eqnarray}}
\def\eea{\end{eqnarray}}
\def\be{\begin{equation}}
\def\ee{\end{equation}}

\newcommand{\roughly}[1]{\mathrel{\raise.3ex\hbox{$#1$\kern-0.85em
\lower1ex\hbox{$\sim$}}}}

\newcommand{\panel}[1]{\includegraphics[width=\textwidth,height=.21\textheight,keepaspectratio]{#1}%
}
\usepackage{dsfont}

\def\Tr{\mathrm{Tr}}

\newcommand{\eq}[1]{eq.~\eqref{#1}}

\numberwithin{equation}{section}
\title{Cosmic Lockdown: When Decoherence Saves the Universe from Tunneling}

\author[a]{Robson Christie,}
\author[a]{Jaewoo Joo,}
\author[b,c]{Greg Kaplanek,}
\author[d]{Vincent Vennin,}
\author[e]{David Wands}

\affiliation[a]{School of Mathematics and Physics, University of Portsmouth, PO1
3FX, United Kingdom}
\affiliation[b]{Department of Electrical Engineering and Computer Science, Syracuse University, NY 13210, USA}
\affiliation[c]{Institute for Quantum \& Information Sciences, Syracuse University, NY 13210, USA}
\affiliation[d]{Laboratoire de Physique de l'Ecole Normale Sup\'erieure, ENS, CNRS, Universit\'e PSL, Sorbonne Universit\'e, Universit\'e Paris Cit\'e, 75005 Paris, France}
\affiliation[e]{Institute of Cosmology \& Gravitation, University of Portsmouth, Dennis Sciama Building, Burnaby Road, Portsmouth, PO1 3FX, United Kingdom}

\emailAdd{robson.christie@port.ac.uk, jaewoo.joo@port.ac.uk, gkaplane@syr.edu, vincent.vennin@phys.ens.fr, david.wands@port.ac.uk}

\date{today}

\begin{document}

\sloppy
\abstract{We investigate how quantum decoherence influences the tunneling dynamics of quantum fields in cosmological spacetimes. Specifically, we study a scalar field in an asymmetric double-well potential during inflation, coupled to environmental degrees of freedom provided by a continuum of spectator fields. This setup enables a systematic derivation of both Markovian and non-Markovian master equations, along with their stochastic unravelings, which we solve numerically. We find that, while decoherence is essential for suppressing quantum interference between vacua, its impact on the relative vacuum populations is limited. Fields heavier than the Hubble scale relax adiabatically toward the true vacuum with high probability, while lighter fields exhibit non-adiabatic enhancements of false-vacuum occupation. Once the system has decohered, quantum tunneling between vacua becomes strongly suppressed, effectively locking the system into the stochastically selected local minimum. This “cosmic lockdown” mechanism is a manifestation of the quantum Zeno effect: environmental monitoring stabilizes enhanced false-vacuum occupation for light fields by preventing them from tunneling.}

 
\maketitle

\section{Introduction}
\label{sec:intro}

The stability of vacuum states in quantum field theory has long been a concern in cosmology. Scalar fields with multiple minima naturally give rise to the possibility of false vacua, in which the universe becomes trapped in a metastable state with relatively higher energy density. The modern understanding of false vacuum decay began with the work of Coleman and Callan \cite{Coleman:1977py,Callan:1977pt}, who showed how tunneling between vacua can be described semi-classically through instanton solutions in field theory. Their formalism was later generalised to gravitational and cosmological settings, first by Coleman and De~Luccia who demonstrated how gravitational effects can modify vacuum decay rates~\cite{Coleman:1980aw}, and later by Hawking and Moss who analysed transitions in de~Sitter space driven by thermal fluctuations over the potential barrier~\cite{Hawking:1981fz}. These results established a physical mechanism that could in principle have serious implications for the large-scale evolution of the universe. Related treatments of barrier crossing within the stochastic-inflation framework have also been explored more recently~\cite{Noorbala:2018zlv,camargo2023phase,rigopoulos2023computing,miyachi2024stochastic}. These stochastic-inflation results differ from ours: here we work at fixed comoving scale (rather than fixed physical scale), but we expect our framework can be adapted to fixed physical regions, and we leave that extension for future work.

A further example comes from the Standard Model itself. The vacuum expectation value of the Higgs field defines the familiar electroweak vacuum, which gives masses to Standard Model particles and determines the structure of low-energy physics. If one assumes that no additional particles or interactions enter the theory up to very high energies, the renormalization-group flow of the Higgs quartic coupling is fixed by measured Standard Model parameters such as the Higgs mass and calculable loop corrections (dominated by the top quark). With these inputs, the Higgs' quartic coupling becomes negative and its effective potential develops a second, deeper minimum at very large field values \cite{Sher:1988mj,Isidori:2001bm, Degrassi:2012ry,Buttazzo:2013uya,Andreassen:2014gha}. In the Standard Model, the associated tunneling rate to the deeper vacuum is typically calculated to be extraordinarily small, so that the electroweak vacuum lifetime vastly exceeds the age of the universe \cite{Isidori:2001bm,Espinosa:2007qp,Chigusa:2018uuj,Markkanen:2018pdo} (see however \cite{Branchina:2013jra}). Nevertheless, it can be unsettling to realize that our universe could, in principle, transition abruptly into a radically different state—one in which the structure of matter and the forces governing it are altered, effectively erasing the world as we know it.

\vspace{2mm}

Reassuringly, the parameters of our universe make such an event extraordinarily unlikely. However, even if the parameters suggested a greater tunneling probability, it would still be important to recognize that these calculations are highly idealized. They are typically performed in Euclidean signature and treat the quantum fields as perfectly isolated, so all evolution occurs simply under a Hamiltonian (and so is unitary). However, in realistic cosmological settings, all fields inevitably interact with a multitude of environmental degrees of freedom, including gravitational perturbations, other matter fields, and possibly hidden sectors. Importantly, in a universe dominated by gravity, its universal coupling ensures that no field is entirely isolated in cosmology. This motivates the central question of this work: how decoherence induced by environmental interactions modifies the standard semiclassical picture of false-vacuum decay and whether it can suppress coherent transitions to influence the fate of cosmological vacua. It also raises the issue of vacuum selection, about which instanton calculations often provide little insight.

\begin{figure}[ht]
\begin{center}
\includegraphics[width=110mm]{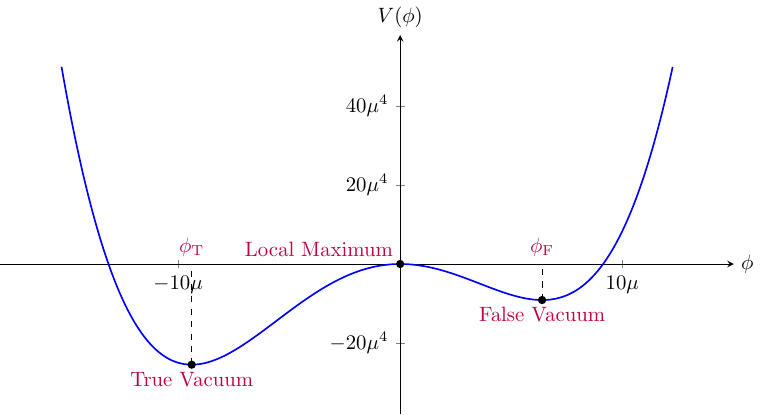}
\caption{\small Potential $V(\phi)$ from \eqref{eq:potential} for the parameter choices $\beta_3 = 0.025$ and  
    $\beta_4 = 0.13$.} \label{fig:potential}
\end{center}
\end{figure}

To address these questions, we instead study here a model of the density matrix for a real scalar field $\phi$ placed in pure de Sitter space, where the line element is
\begin{eqnarray}
\exd s^2 = - \exd t^2 + a^2(t) \exd \mathbf{x}^2 \qquad \mathrm{with\ scale\ factor\ } a(t) = e^{Ht} 
\end{eqnarray}
where $H$ is the Hubble scale and cosmic time $t$ is related to the number of $e$-folds through
\begin{equation} \label{efolds}
N = Ht \ .
\end{equation}
The main feature of our setup is that the field evolves in a potential \(V(\phi)\) with two inequivalent vacua while additionally interacting with an environment, allowing us to capture the effects of decoherence on vacuum dynamics. Throughout we model the environment as a continuum of massive spectator scalars \(X^{\mfa}\) coupled locally to \(\phi\) via quartic interactions of the form \(\phi^{4-k} (X^{\mfa})^{k}\) with \(k = 1,2,3\); in the main text we focus on a \(\phi (X^{\mfa})^{3}\) coupling, while the \(\phi^{3} X^{\mfa}\) and \(\phi^{2} (X^{\mfa})^{2}\) cases and their corresponding master equations are discussed in App.~\ref{app:derivation}. Throughout we consider the quartic potential
\begin{equation}
V(\phi)
= -\frac{1}{2}\mu^2 \phi^2
  + \frac{2}{3}\,\beta_3 \mu \phi^3
  + \frac{1}{4}\,(\beta_4^{2} - \beta_3^{2})\,\phi^4,
\label{eq:potential}
\end{equation}
see Fig.~\ref{fig:potential}, with
\begin{equation}
\mu>0,\qquad \beta_4>0,\qquad \beta_4^{2}>\beta_3^{2},
\label{eq:bounded}
\end{equation}
so that the potential is bounded from below. This quartic admits a local maximum at the origin,
\begin{equation}
\phi_{\mathrm{M}} = 0,
\end{equation}
and two nondegenerate minima at
\begin{equation}
\phi_{\mathrm{T}} = - \frac{\mu}{\beta_4 - \beta_3},
\qquad
\phi_{\mathrm{F}} = + \frac{\mu}{\beta_4 + \beta_3}.
\label{eq:minima}
\end{equation}
Evaluating the potential at the minima gives
\begin{equation}
V(\phi_{\mathrm{T}})
= - \frac{3 \beta_4 - \beta_3}{12 (\beta_4 - \beta_3)^3}\,\mu^4,
\qquad
V(\phi_{\mathrm{F}})
= - \frac{3 \beta_4 + \beta_3}{12 (\beta_4 + \beta_3)^3}\,\mu^4,
\label{eq:vacua-values}
\end{equation}
so the vacuum-energy splitting is
\begin{equation}
\Delta V \equiv V(\phi_{\mathrm{T}}) - V(\phi_{\mathrm{F}})
= \frac{4 \beta_3 \beta_4^{3}}{3 (\beta_4^{2} - \beta_3^{2})^{3}}\,\mu^{4}.
\label{eq:deltaV}
\end{equation}
For the parameter range \eqref{eq:bounded}, \(\Delta V>0\), hence \(\phi_{\mathrm{T}}\) is the true vacuum and \(\phi_{\mathrm{F}}\) is the false vacuum in the regime of interest. The curvature at the barrier is fixed by
\begin{equation}
V''(\phi_{\mathrm{M}}) = -\mu^{2},
\label{eq:barrier}
\end{equation}
so \(\mu\) directly quantifies the instability scale. The effective masses governing small fluctuations about the two vacua follow from \eqref{eq:potential} and \eqref{eq:minima}:
\begin{equation}
m^{2}_{\mathrm{F}} = V''(\phi_{\mathrm{F}})
= \mu^{2} \left( 1 + \frac{\beta_4 + \beta_3}{\beta_4 - \beta_3} \right),
\qquad
m^{2}_{\mathrm{T}} = V''(\phi_{\mathrm{T}})
= \mu^{2} \left( 1 + \frac{\beta_4 - \beta_3}{\beta_4 + \beta_3} \right).
\label{eq:masses}
\end{equation}
For \(\beta_4^{2}>\beta_3^{2}\) one has
\begin{equation}
m^{2}_{\mathrm{T}} > m^{2}_{\mathrm{F}},
\end{equation}
so the true vacuum has a larger effective mass. In the near-symmetric limit \(|\beta_3|\ll \beta_4\),
\begin{equation}
\Delta V
\simeq \frac{4 \beta_3}{3 \beta_4^{3}}\,\mu^{4}\left[1
+ \mathcal{O}\!\left(\frac{\beta_3}{\beta_4}\right)^{2}\right],
\end{equation}
and the two vacua become nearly degenerate, which is useful for testing environment-induced decoherence in a controlled small-splitting regime.
We can see both effective masses are proportional to the mass scale of the potential $\mu$. To quantify adiabaticity we define the dimensionless parameter
\begin{equation}
  \tilde \mu \equiv \frac{\mu}{H}.
\end{equation}
Physically $\tilde \mu$ compares the growth rate of the effective inverted oscillator at the barrier top to the Hubble rate (see Appendix~\ref{app:adiabatic} for related illustrative example). We also couple to other spectator fields, schematically represented by additional scalar fields $\chi$ via four-point interactions of the form
\begin{equation}
\mathcal{L}_{\mathrm{int}} = \underline{\lambda}_1 \phi^3 \chi + \underline{\lambda}_2 \phi^2 \chi^2 + \underline{\lambda}_3 \phi \chi^3  \qquad \mathrm{with} \ \underline{\lambda}_j > 0 \ .
\end{equation}
Here $ \underline{\lambda}_j$ are bare dimensionless couplings in the action (whereas  renormalized couplings $\lambda_j$ appear in the dynamical equations instead). Our focus in the main text will be on the third $\phi\chi^3$ interaction, although we explore  all interactions in Appendix \ref{app:derivation}.

All fields are treated as spectators in de Sitter space, so backreaction on the geometry is neglected.  To render the dynamics tractable while retaining the essential physics of the false-vacuum system, we perform a coarse-graining over a fixed comoving box and neglect gradient energy within the patch (see \S\ref{subsec:fourpoint_Lindblad_SSE} for details).

Our setup allows us to derive both Markovian and non-Markovian master equations that describe the reduced dynamics of the scalar field in a quartic potential. Since analytic solutions of these equations, even just for the coarse-grained field, are intractable, we solve them numerically using a truncated Fock basis to explore how environmental interactions influence vacuum selection in an inflationary background. Appendix~\ref{app:adiabatic} provides an illustrative and analytically tractable quadratic parametric oscillator example. These master equations provide an ensemble-averaged description of the reduced density matrix, which captures the coarse-grained evolution in phase space, represented by the Wigner function. This is contrasted in our results with the dynamics of individual stochastic trajectories, which can exhibit more varied behavior. In this sense, master equations, in particular Lindblad, offer a natural description of vacuum selection under decoherence at the level of averaged, physically observable quantities.

\vspace{2mm}

Our analysis reveals several key results:

\begin{itemize}
    
\item First, the adiabaticity parameter $\tilde{\mu}$ controls the relative vacuum populations. If the Hubble rate is small compared to the mass scale set by the potential, the evolution is adiabatic: the field follows the instantaneous ground state and becomes localized in the true vacuum at late times. By contrast, when $\tilde{\mu}$ is sufficiently small, the evolution is rapid and non-adiabatic, and the system relaxes to a mixed state with comparable populations in both minima, with a non-adiabatic enhancement of the false-vacuum occupation.

\item Second, we find that decoherence itself does not have much influence on {\it which} vacuum is chosen. Introducing a nonzero coupling to the environment ($\underline{\lambda} \neq 0$) does not significantly affect the relative probability of ending in the true or false vacuum. Instead, the principal role of decoherence is to suppress quantum interference between vacua, ensuring that once a vacuum is selected the system cannot tunnel through the barrier. 

\item Third, the time-dependence of the de Sitter background means that the potential acquires a growing prefactor with time (coming from $\sqrt{-g}$ required on the grounds of covariance), which increases the energetic separation between minima and sharpens the wells in phase space at late times. This stretching of the potential in phase space helps to reinforce the outcome of vacuum selection.

 \item Fourth, at late times, once the system has decohered into a definite local minimum, quantum tunneling between vacua is strongly suppressed. Even if the field settles into a false vacuum, continuous environmental “monitoring’’ or “measurement’’ induces a quantum Zeno effect~\cite{itano1990quantum,li2021quantum}, suppressing any subsequent tunneling back to the true vacuum. In this sense, decoherence gives rise to something we call a {\it cosmic lockdown} mechanism, stabilizing whichever vacuum the system has reached and preventing further evolution through tunneling.
 
\end{itemize}

These findings build on earlier work showing that scalars, gravitons and other environmental degrees of freedom induce decoherence in cosmology \cite{Sakagami:1987mp,Brandenberger:1990bx,Matacz:1992mk,Lombardo:1995fg,Calzetta:1995ys,Polarski:1995jg,Kiefer:1998qe,Lombardo:2005iz,Burgess:2006jn,Prokopec:2006fc,Sharman:2007gi,Kiefer:2008ku,Kiefer:2010pb,Bachlechner:2012dg,Franco:2011fg,Burgess:2014eoa,Nelson:2016kjm,boddy2017decoherence,Bao:2019ghe,Brahma:2020zpk,Colas:2022hlq,Burgess:2022nwu,DaddiHammou:2022itk,Colas:2022kfu,Sou:2022nsd,Boutivas:2023mfg,Colas:2024xjy,Colas:2024ysu,deKruijf:2024ufs,Burgess:2024eng,Burgess:2025dwm,Cespedes:2025zqp,deKruijf:2025jya,Sano:2025ird,Lopez:2025arw,Takeda:2025cye} which are more closely related to semi-classical instanton studies first mentioned. Our minimal toy model shows how environmental interactions can alter the standard picture of false-vacuum decay. Unlike \cite{Bachlechner:2012dg}, which assumes the system starts in the false vacuum, we consider an initial state in the instantaneous ground state, which is delocalized across both vacua.

\section{Coarse-grained Schr\"odinger equation }
Before incorporating the effects of decoherence, we derive in this section the effective Hamiltonian that governs the closed-system dynamics of our scalar field. The system $\mathcal{S}$ we consider is a real scalar field $\Phi$ evolving in the asymmetric potential $V(\Phi)$ given in Eq.~\eqref{eq:potential}. The corresponding continuum action in a de Sitter background is
\begin{equation} \label{eq:actionsystem}
S_{\mathcal{S}} = \int \exd t   a^{3} \int \exd^3 \mathbf{x}  \left[ \frac{1}{2} \dot{\Phi}^2 - \frac{1}{2a^2} | \boldsymbol{\nabla} \Phi |^2 - V(\Phi) \right] .
\end{equation}
The corresponding Hamiltonian is given by 
\begin{equation}
    H_{\mathcal{S}}=\int \exd^3 x \left[\frac{\Pi^2}{2 a^3}+\frac{a}{2}|\nabla \Phi|^2+a^3 V(\Phi)\right]\quad \text{with} \quad \dot{\Phi}=\frac{\Pi}{a^3}.
\end{equation}
To simplify the resulting dynamics, we coarse grain this Hamiltonian over a region $R$ with constant comoving volume
\begin{equation} \label{eq:vol}
    \mathtt{vol} \equiv \int_{R} \mathrm{d}^3x.
\end{equation}
The time-dependent physical volume of this region is $a^{3}(t)\,\mathtt{vol}$. We define the canonical coarse-grained operators for the system field and momentum as
\begin{equation} \label{eq:cannonical_pair}
    \phi(t)\equiv \frac{1}{\mathtt{vol}}\int_{R} \mathrm{d}^{3}x\;\Phi(\bm x,t),
    \qquad
    \pi_\phi(t)\equiv \int_{R} \mathrm{d}^{3}x\; \Pi_\Phi(\bm x,t) = \int_{R} \mathrm{d}^{3}x\; a^3 \dot\Phi(\bm x,t).
\end{equation}
From the canonical commutation relations of the microscopic field, $[\hat\Phi(\bm x,t),\hat\Pi_\Phi(\bm y,t)]=i\delta^{(3)}(\bm x-\bm y)$, it follows that
\begin{equation}
    [\hat \phi(t),\hat \pi_\phi(t)] = \frac{1}{\mathtt{vol}}\int_{R} \mathrm{d}^{3}x\int_{R} \mathrm{d}^{3}y\; [\hat\Phi(\bm x,t),\hat\Pi_\Phi(\bm y,t)] = i
\end{equation}
with the average field density inside the comoving box canonically conjugate to the total field momentum contained within the box.

By assuming the modes are approximately homogeneous within the box, we may neglect the gradient energy contributions. The Hamiltonian simplifies to
\begin{equation}
H_S  \simeq  \mathtt{vol}\left[\frac{\Pi^2}{2 a^3}+a^3 V(\Phi)\right] .\label{eq:action_box_system}
\end{equation}
Changing time parametrization from cosmic time $t$, to $e$-folds $N$ following Eq.~\eqref{efolds} and writing the Hamiltonian in terms of the canonical pair of coarse-grained operators defined in Eq.~\eqref{eq:cannonical_pair} yields
\begin{equation} \label{eq:effHamiltonian}
\hat K_S(N)
= \frac{e^{-3N}}{2H\mathtt{vol}}\,\hat \pi_\phi^{2}
+ \frac{e^{3N} \mathtt{vol}}{H}\,
\left[ -\frac{1}{2}\mu^2 \hat \phi^2
  + \frac{2}{3}\,\beta_3 \mu \hat \phi^3
  + \frac{1}{4}\,(\beta_4^{2} - \beta_3^{2})\,\hat \phi^4\right],
\end{equation}
Across all simulations we fix
\begin{equation} \label{eq:well-params}
    \mu = 0.5 \;[{\rm mass}^{+1}],\qquad
    \beta_3 = 0.025,\qquad
    \beta_4 = 0.13,\qquad
    \mathtt{vol}=4\sqrt{2}\simeq 5.66\; [{\rm mass}^{-3}] \, .
\end{equation}
These numerical values are arbitrary and chosen for convenience.  In
particular, we choose $\mathtt{vol}$ such that the kinetic prefactor and the characteristic potential prefactor become comparable near the conventional
origin $N=0$.  Equating these characteristic scales,
\begin{equation}
\frac{e^{-3N}}{2H\mathtt{vol}}\;\mu^{-2}
\;\sim\;
\frac{e^{3N}\mathtt{vol}}{H}\;\mu^{4},
\end{equation}
gives the crossover $e$-fold
\begin{equation}
N_\star
= \frac16\ln\!\left[\frac{1}{2\,\mathtt{vol}^{2}\mu^{6}}\right]. \label{eq:crossover}
\end{equation}
Since the overall normalization of the scale factor is arbitrary, one may equivalently shift the $e$-fold origin so that $N_\star=0$ (i.e.\ $a=1$ at crossover), in which case the physical coarse-graining volume at crossover is $V_{\rm phys}(N_\star)=a^3(N_\star)\,\mathtt{vol}=\mathtt{vol}$ (more generally $V_{\rm phys}(N)=a^3(N)\,\mathtt{vol}$).

To simulate the coarse-grained dynamics, we assume the system initially occupies the ground state of the effective Hamiltonian in Eq.~\eqref{eq:effHamiltonian} in the distant past. We then evolve the state forward in time using the Schr\"odinger equation,
\begin{equation}
\partial_N \ket{\psi}
= -i \hat K_S(N)\ket{\psi} ,
\label{eq:Schroedinger_N}
\end{equation}
which can be equivalently expressed for the density operator $\hat\rho = \ket{\psi}\!\bra{\psi}$ via the von~Neumann equation,
\begin{equation}
\partial_N \hat\rho
= -i\bigl[\hat K_S(N),\hat\rho\bigr] .
\label{eq:vonNeumann_N}
\end{equation}

In the next section, we introduce the cosmological master equations and stochastic unravelings used throughout the rest of the paper.

\section{Markovian master equations}

To incorporate decoherence from a memoryless environment, we first introduce the general forms of the open-system dynamical equations used throughout this work. Specifically, the generalisation of the closed-system von~Neumann equation~\eqref{eq:vonNeumann_N} to open systems is a Gorini-Kossakowski-Lindblad-Sudarshan (GKLS)~\cite{lindblad1976generators,gorini1976completely} master equation of the form
\begin{equation}
\partial_N\hat\rho
= -i\bigl[\hat K_S(N),\hat\rho\bigr]
- \frac12 \sum_\alpha \gamma_\alpha(N)\bigl[\hat L_\alpha,\bigl[ \hat L_\alpha, \hat\rho\bigr]\bigr] ,
\label{eq:GKLS_general}
\end{equation}
where the $\hat L_\alpha$ are Hermitian Lindblad operators and $\gamma_\alpha(N)\ge0$ are time-dependent decoherence rates.

Correspondingly, we introduce the general form of the normalized It\^o stochastic Schr\"odinger equation (SSE)~\cite{belavkin1989nondemolition,percival} that unravels~\eqref{eq:GKLS_general}, which can be written as
\begin{multline}
    \qquad \qquad \dd\ket{\psi}=
 -i\hat K_S(N)\ket{\psi}\dd N
- \frac12\sum_\alpha \gamma_\alpha(N)\Bigl(\hat L_\alpha- \ev*{\hat L_\alpha}\Bigr)^{2}\ket{\psi}\dd N \\
+ \sum_\alpha \sqrt{\gamma_\alpha(N)}\bigl(\hat L_\alpha - \ev*{\hat L_\alpha}\bigr)\ket{\psi}\; \dd W_\alpha ,
\label{eq:SSE_general}
\end{multline}
where the $\dd W_\alpha$ are independent real Wiener increments with $\mathbb E[\dd W_\alpha]=0$ and $\mathbb E[\dd W_\alpha \dd W_\beta]=\dd N \delta_{\alpha\beta}$. The SSE~\eqref{eq:SSE_general} is a nonlinear, norm-preserving equation in which the noise couples through the fluctuation operators $\hat L_\alpha - \ev*{\hat L_\alpha}$. It corresponds to an effective environmental monitoring with a system-bath interaction Hamiltonian
\begin{equation}
\hat H_{\rm int} = \sum_{\alpha} \hat L_{\alpha}\otimes \hat B_{\alpha} ,
\end{equation}
together with Markovian interaction-picture bath correlators
\begin{equation}
\ev*{\hat B_{\alpha_1}(t)\hat B_{\alpha_2}(t')}
= \delta_{\alpha_1\alpha_2}\,\gamma_{\alpha_1}(N)\,\delta(t-t') ,
\end{equation}
which encode the decoherence rates $\gamma_\alpha(N)\ge 0$. In this picture, Eq.~\eqref{eq:SSE_general} describes continuous measurement or bath-induced localization at the level of individual trajectories~\cite{petruccione}.

\subsection{Four-point system-bath interaction master equations}
\label{subsec:fourpoint_Lindblad_SSE}

As discussed in \S\ref{sec:intro}, we study a de Sitter analogue of the Caldeira-Leggett model~\cite{caldeira1983path,petruccione}, in which a system $\mathcal{S}$ interacts with an environment $\mathcal{E}$ through an overall action of the form
\begin{eqnarray} \label{fullaction}
S = S_{\mathcal{S}} + S_{\mathcal{E}} + S_{\mathrm{int}} \, .
\end{eqnarray}
In addition to the system action given previously in Eq.~\eqref{eq:actionsystem}, we also consider a family of massive real scalar fields $\{ X^{\mfa} \}$ each with their own distinct mass $m_{\mfa}$ such that
\begin{equation}
S_{\mathcal{E}} =  \sum_{\mfa} \int \exd t \;  a^{3} \int \exd^3 \mathbf{x} \; \bigg[ \frac{1}{2} ( \dot{X}^{\mfa} )^2 - \frac{1}{2a^2} | \boldsymbol{\nabla} X^{\mfa} |^2 -  \frac{1}{2} m_{\mfa}^2 (X^{\mfa})^2  \bigg] \ .
\end{equation}
Finally, we assume that the system field $\Phi$ linearly couples to all environmental fields $X^{\mfa}$ through an interaction of the form
\begin{equation}
S_{\mathrm{int}} = - \underline{\lambda} \sum_{\mfa}   g_{\mfa}\int \exd t \; a^{3} \int \exd^3 \mathbf{x} \; \Phi (X^{\mfa})^3
\end{equation}
with distinct dimensionless couplings $\{ g_{\mfa} \}$ and another bare dimensionless coupling $\underline{\lambda}$ used for later bookkeeping. In Appendix \ref{app:derivation} we explore more general quartic (four-point) interactions between system and environment, see Eq.~\eqref{eq:quartic_interactions}. 

We next assume both the $\phi$ and $\chi$ fields are approximately homogeneous within the box and neglect the gradient energy contributions. The total system environment action simplifies to
\begin{equation}
S  \simeq  \mathtt{vol} \int \exd t \;  a^{3} \bigg\lbrace \frac{\dot{\phi}^2 }{2 } - V\left( \phi \right) +  \sum_{\mfa} \bigg[ \frac{ (\dot{\chi}^{\mfa})^2  - m_{\mfa}^2 (\chi^{\mfa})^2}{2} - \underline{\lambda} \, g_{\mfa} \phi (\chi^{\mfa})^3 \bigg]  \bigg\rbrace \label{eq:action_box}
\end{equation}
where we have similarly defined the coarse-grained environmental variables
\begin{equation}
\chi^{\mfa}(t) \equiv \frac{1}{\mathtt{vol}}\int_{R} \mathrm{d}^{3}x\; X^{\mfa}(\bm x,t) \ .
\end{equation}
Next, we assume the environment consists of a continuous spectrum of fields rather than a discrete set. Taking the continuum limit for their densely distributed masses ($m_{\mfa} \to m$), we replace the discrete sum over couplings with an integral:
\begin{equation}
\sum_{\mfa} g_{\mfa} \to \frac{1}{\mu} \int_0^\infty \exd m\, G(m) ,
\end{equation}
where $\mu$ is the reference mass scale from the potential defined in \eqref{eq:potential}. As the frequencies are continuous we replace the label on the fields such that $\chi^{\mfa} \to \chi(m)$ and the couplings become $g_{\mfa} \to G(m)$, giving
\begin{equation}
S  \simeq \mathtt{vol} \int \exd t \;  a^{3} \bigg\{ \frac{\dot{\phi}^2}{2 }  - V\left(\phi \right) +  \int_0^\infty \frac{ \exd m }{\mu} \; \bigg[ \frac{\dot{\chi}^2(m) - m^2 \chi^2(m)}{2 }  - \underline{\lambda} \, G(m) \,\phi \chi^3(m) \bigg]  \bigg\} .
\end{equation}
We choose the spectral density $G(m)$ to be
\begin{equation}
G(m)
=\left( \frac{m}{\mu} \right)^{\frac{3}{2}} e^{-\tfrac{3m}{2\Lambda}},
\label{eq:J_ohmic}
\end{equation}
which compensates the mass scaling of the correlators of the homogeneous bath modes $\chi(m)$ and is precisely the choice that makes the $\Phi X^{3}$ channel yield a time-local (Markovian) master equation; see Appendix~\ref{app:derivation} for the explicit calculation. The cutoff $\Lambda$ is physically reasonable, since the system should only couple efficiently to environmental fields of comparable mass, while very heavy modes ($m\gg\Lambda$) can be integrated out in the Wilsonian sense and simply renormalise the self-interaction parameters in $V(\phi)$. In Appendix~\ref{app:derivation} we also show how starting from a single massive environmental scalar with generic quartic couplings $\Phi^{4-k} X^{k}$ and promoting $X$ to a continuum weighted by suitable powers of the same spectral density leads to Markovian master equations for $k=1,2,3$, each with a different Lindblad structure.

The total Lagrangian $L$ is defined by $S = \int \exd t \, L$, which gives
\begin{equation}
  L(t) =
  \mathtt{vol} \cdot a^{3} \Biggl\{
      \frac{1}{2} \dot{\phi}^{2}-V(\phi)
     - \frac{1}{\mu}\int_{0}^{\infty} \dd m\,
       \Bigl[ \frac{1}{2} \dot{\chi}_m^{2} 
          -\frac12 m^{2}\chi_m^{2} + \underline{\lambda} \, G(m) \, \phi \chi^3_m        
       \Bigr]
     \Biggr\} ,
  \label{eq:L_ohmic}
\end{equation}
where the homogeneous box-averaged mode \(\phi(N)\) serves as the system (with dimensions of $\mathrm{mass}^{+1}$), while the continuum of heavy spectator modes \(\{ \chi_m(N) \}\) acts as the environment (also with dimensions of $\mathrm{mass}^{+1}$ in our conventions). We assume that these spectator modes begin in their ground state in the infinite past. The potential $V(\phi)$ in \eqref{eq:potential} is asymmetric and always admits a false vacuum, as shown in Fig.~\ref{fig:potential}.

In the main text we specialise to the $\phi\chi^{3}$ coupling, which, after renormalisation of the bare coupling, yields a purely decohering GKLS master equation with Lindblad operator proportional to $\hat\phi$; see Appendix~\ref{app:derivation} for details. Within the general notation of \eqref{eq:GKLS_general} this corresponds to a master equation of the form
\begin{equation}
\partial_N \hat\rho
= -\,i\bigl[\hat K_S(N),\hat\rho\bigr]
 -\frac{131\pi\lambda^{2} e^{6N}}{512\,\mu^{5}\,\mathtt{vol}}\,[\hat\phi,[\hat\phi,\hat\rho]] . \label{eq:GKLS_phi_only}
\end{equation}
A convenient normalised nonlinear SSE of the type \eqref{eq:SSE_general} that unravels \eqref{eq:GKLS_phi_only} is
\begin{equation}
\dd \ket{\psi}
=- i\hat K_S\ket{\psi} \dd N
-\frac{131\pi\lambda^{2} e^{6N}}{512\,\mu^{5}\,\mathtt{vol}}\Big(  \hat \phi-\langle \hat \phi\rangle \Big)^2\ket{\psi}\dd N
+\sqrt{\frac{131\pi\lambda^{2} e^{6N}}{256\,\mu^{5}\,\mathtt{vol}}}\big(\hat \phi-\langle \hat \phi \rangle\big)\ket{\psi}\; \dd W,
\label{eq:SSE_phi_only}
\end{equation}
where $\dd W$ is a real Wiener increment. Ensemble averages of these pure state trajectories over noise realisations yield the solution of the $\hat\phi$-Lindblad equation \eqref{eq:GKLS_phi_only}. This dynamics describes pure dephasing and localization in the $\hat\phi$ basis induced by the $\chi_m$ environment.

\section{Numerical results: from kinetic to potential dominance}
\label{sec:results}
\FloatBarrier

\begin{figure}
    \centering
    \begin{subfigure}[t]{0.32\textwidth}\centering
        \panel{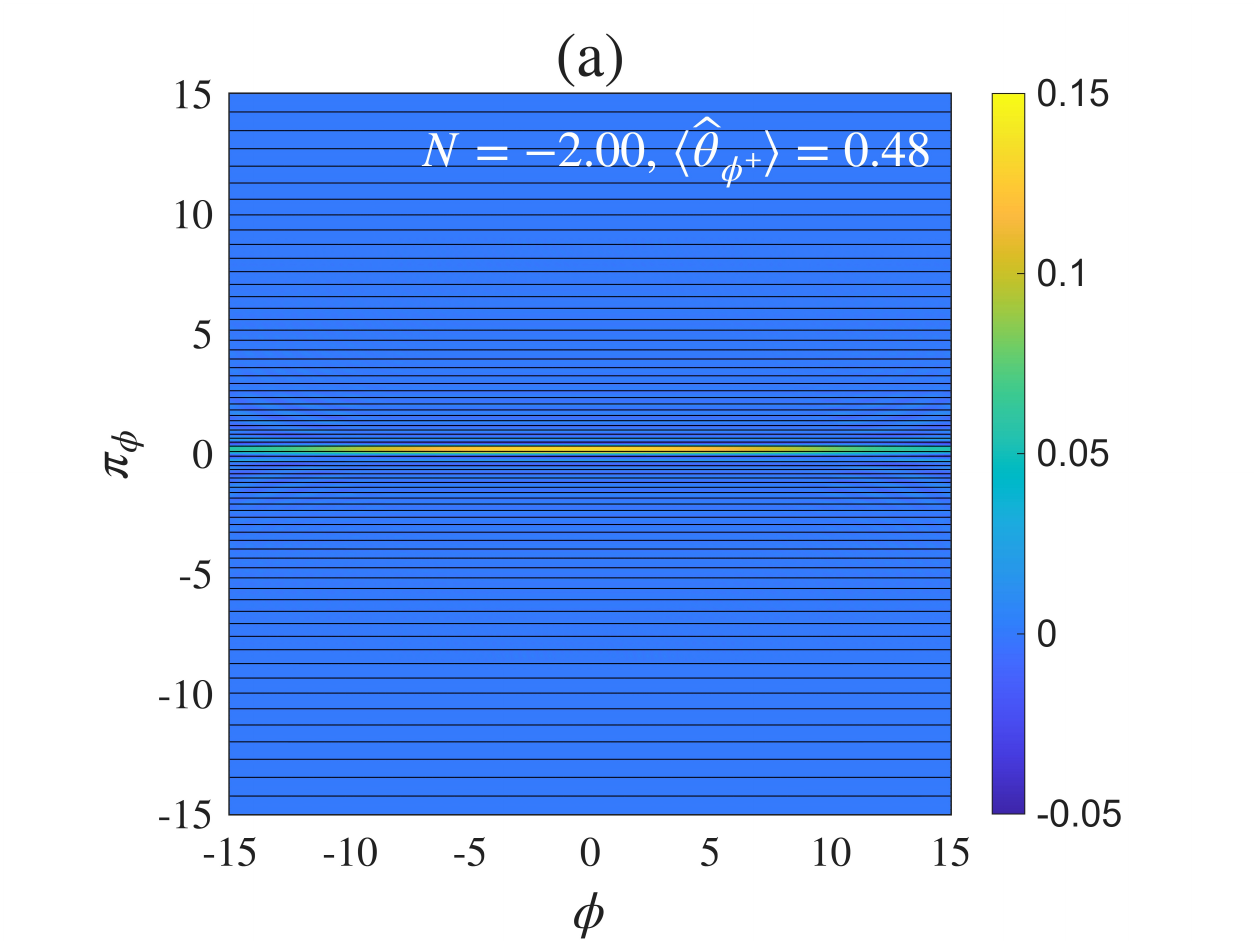}
    \end{subfigure}\hfill
    \begin{subfigure}[t]{0.32\textwidth}\centering
        \panel{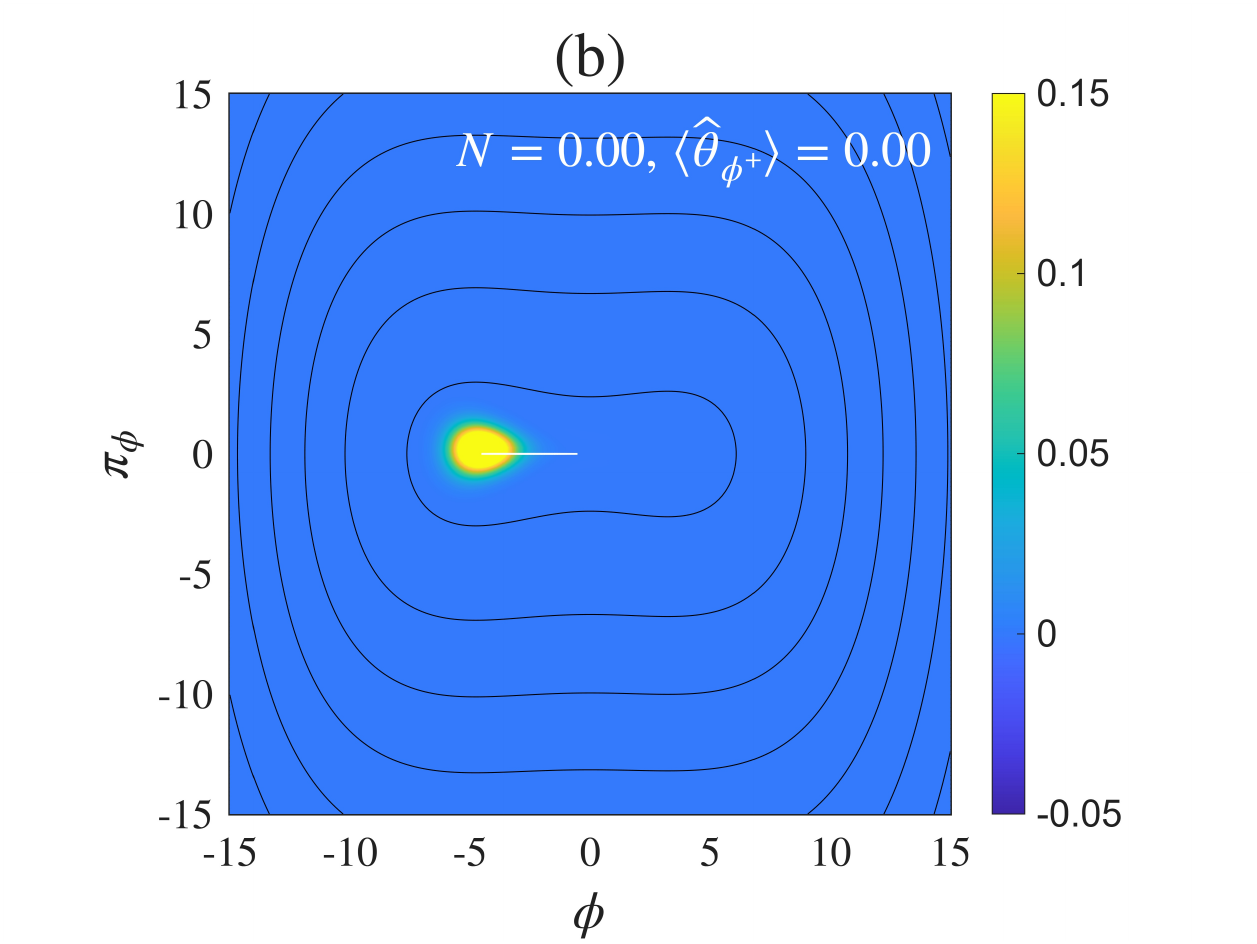}
    \end{subfigure}\hfill
    \begin{subfigure}[t]{0.32\textwidth}\centering
        \panel{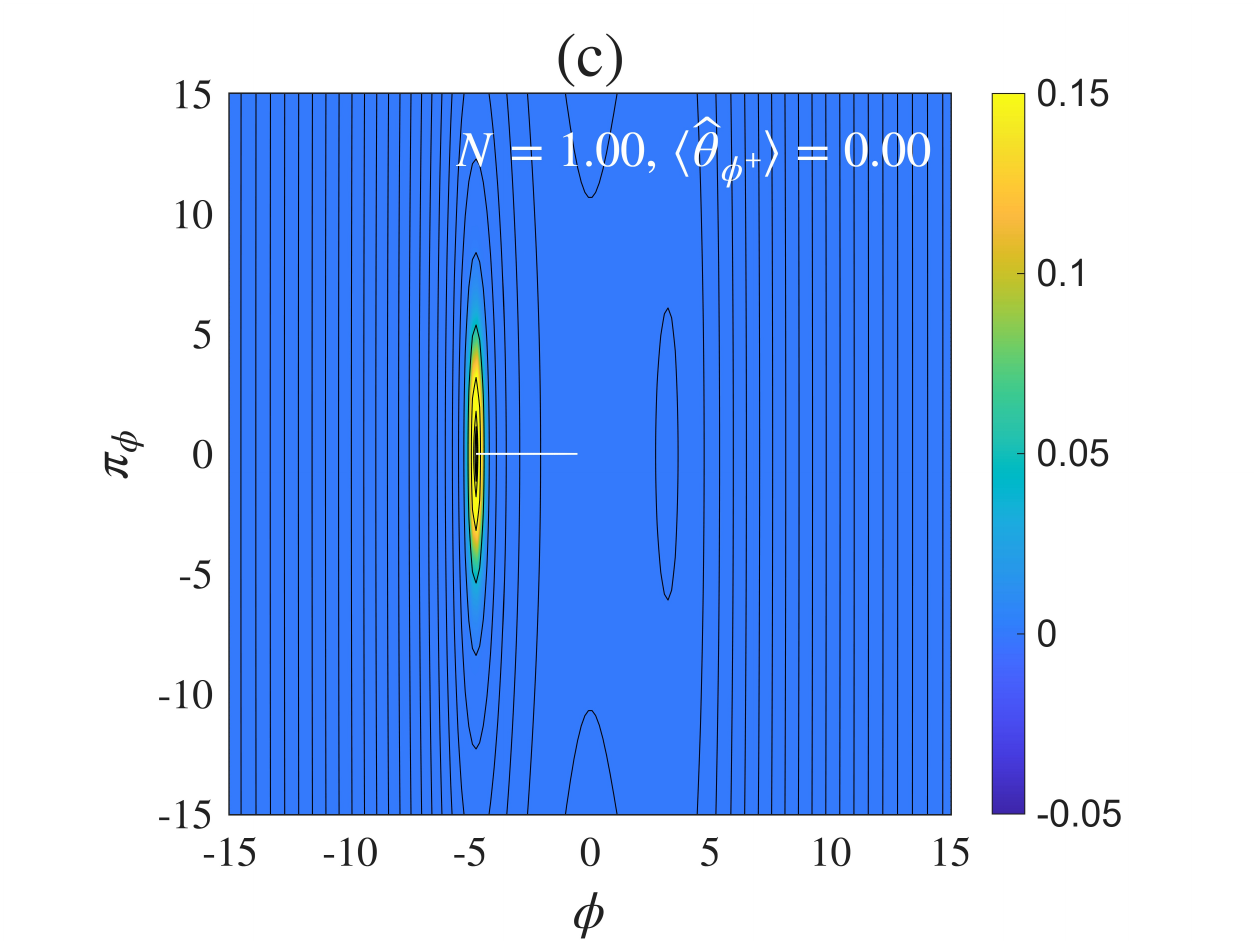}
    \end{subfigure}
    \caption{\small
      {\bf Adiabatic limit} ($\tilde \mu \to \infty$). Wigner functions showing the instantaneous Hamiltonian~\eqref{eq:effHamiltonian} ground state at three snapshot times. Panel~(a) is the instantaneous ground state at e-fold $N=-2$, which is the initial state used in every simulation in this paper. The black curves are equal-energy contours of $K_S(N)$. Axes are $\phi$ and $\pi_\phi$. In-panel text reports the $e$-fold $N$ and the false-vacuum projector expectation value 
    $\langle \hat\theta_{\phi^{+}}\rangle = \mathrm{Tr}\big[\hat\theta_{\phi^{+}}\hat\rho(N)\big]$. The white line corresponds to the evolution prior to the snapshot  of the phase-space expectation values $\ev*{\hat \phi}$ and $\ev*{\hat \pi_{\phi}}$. The sonified video corresponding to these plots is available
    \href{https://youtube.com/shorts/mCOya84QF-g}{via this link} with the sonification method described in \cite{christie2024sound}. }
    \label{fig:WignerAdiabatic}
\end{figure}

\begin{figure}
    \centering
    \begin{subfigure}[t]{0.32\textwidth}\centering
        \panel{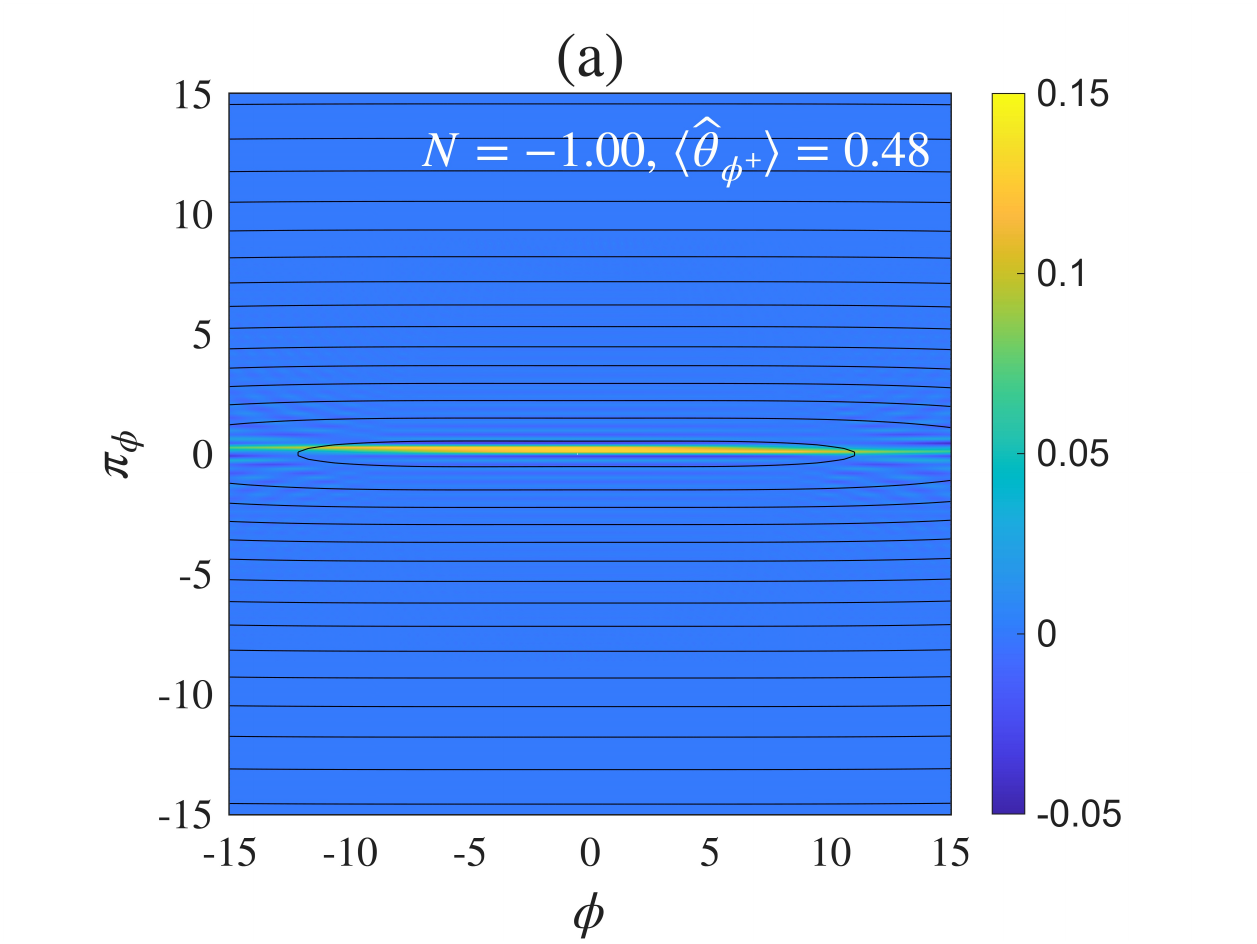}
    \end{subfigure}\hfill
    \begin{subfigure}[t]{0.32\textwidth}\centering
        \panel{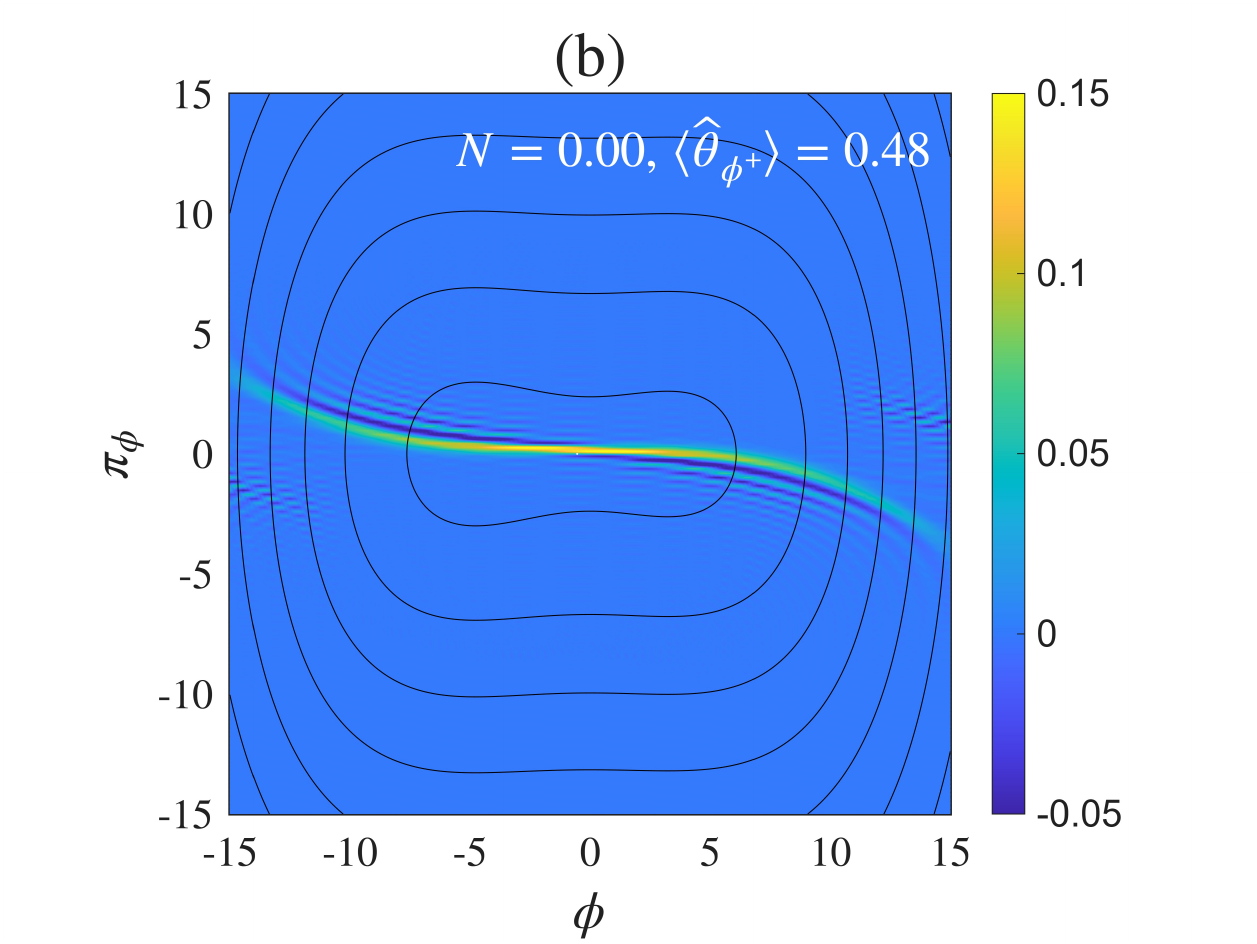}
    \end{subfigure}\hfill
    \begin{subfigure}[t]{0.32\textwidth}\centering
        \panel{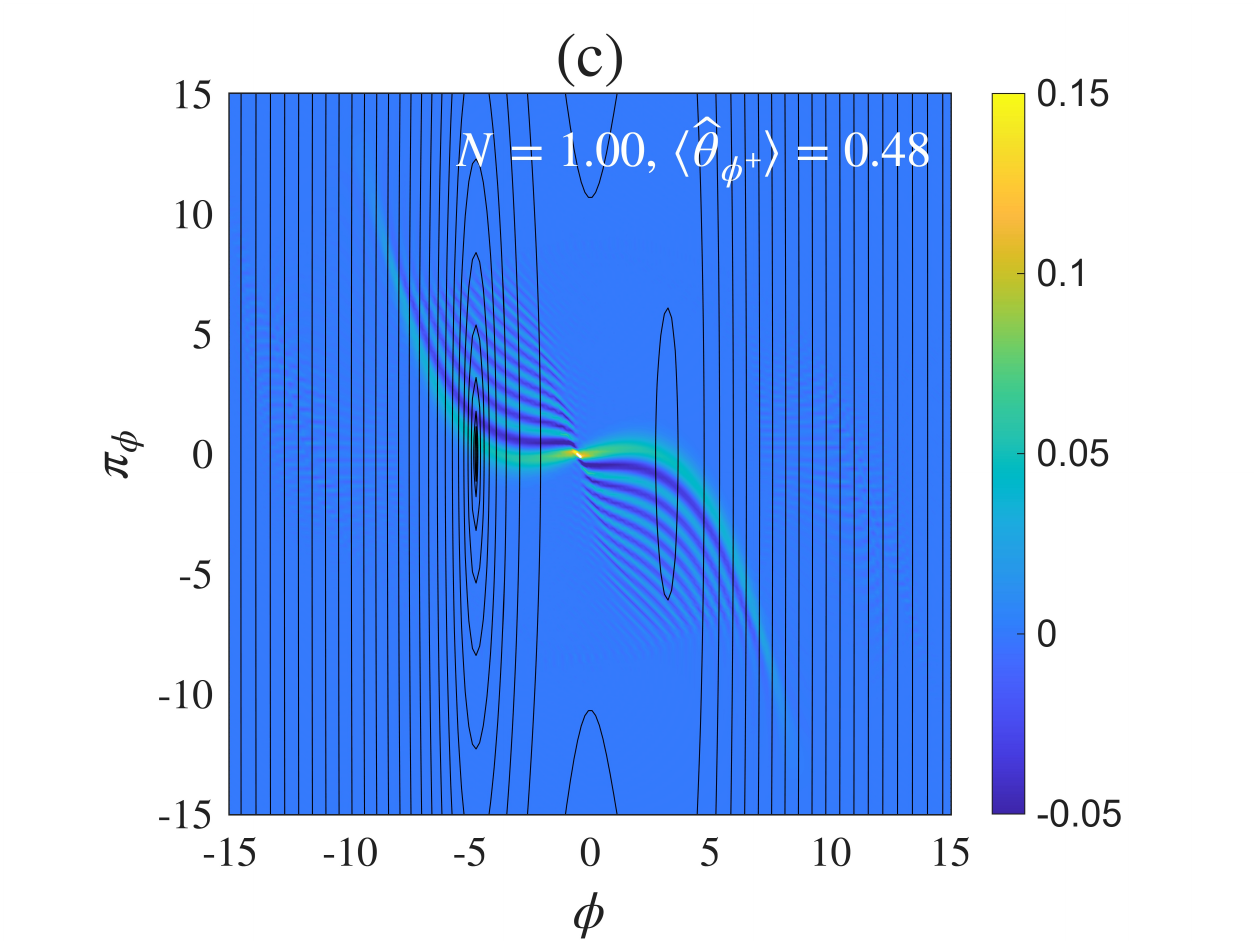}
    \end{subfigure}
    \caption{\small
    {\bf Schr\"odinger dynamics}. Wigner functions for Hamiltonian evolution~\eqref{eq:Schroedinger_N} without decoherence.
    The system is evolved unitarily from the instantaneous ground state at $N=-2$. The sonified video corresponding to these plots is available
    \href{https://youtube.com/shorts/1sZAcApsoVs}{via this link}. }
    \label{fig:WignerHam}
\end{figure}

\begin{figure}
    \centering
    \begin{subfigure}[t]{0.32\textwidth}\centering
        \panel{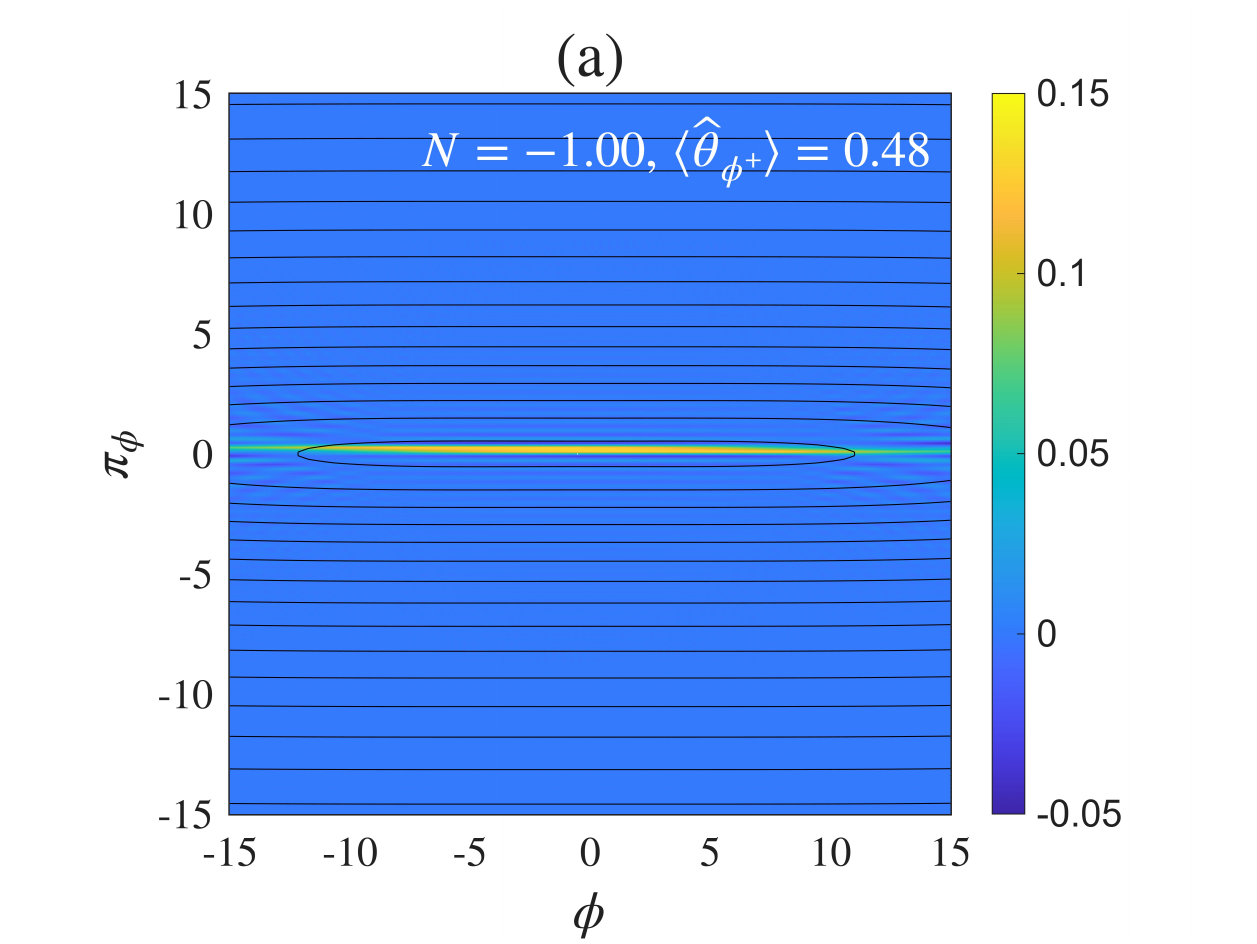}
    \end{subfigure}\hfill
    \begin{subfigure}[t]{0.32\textwidth}\centering
        \panel{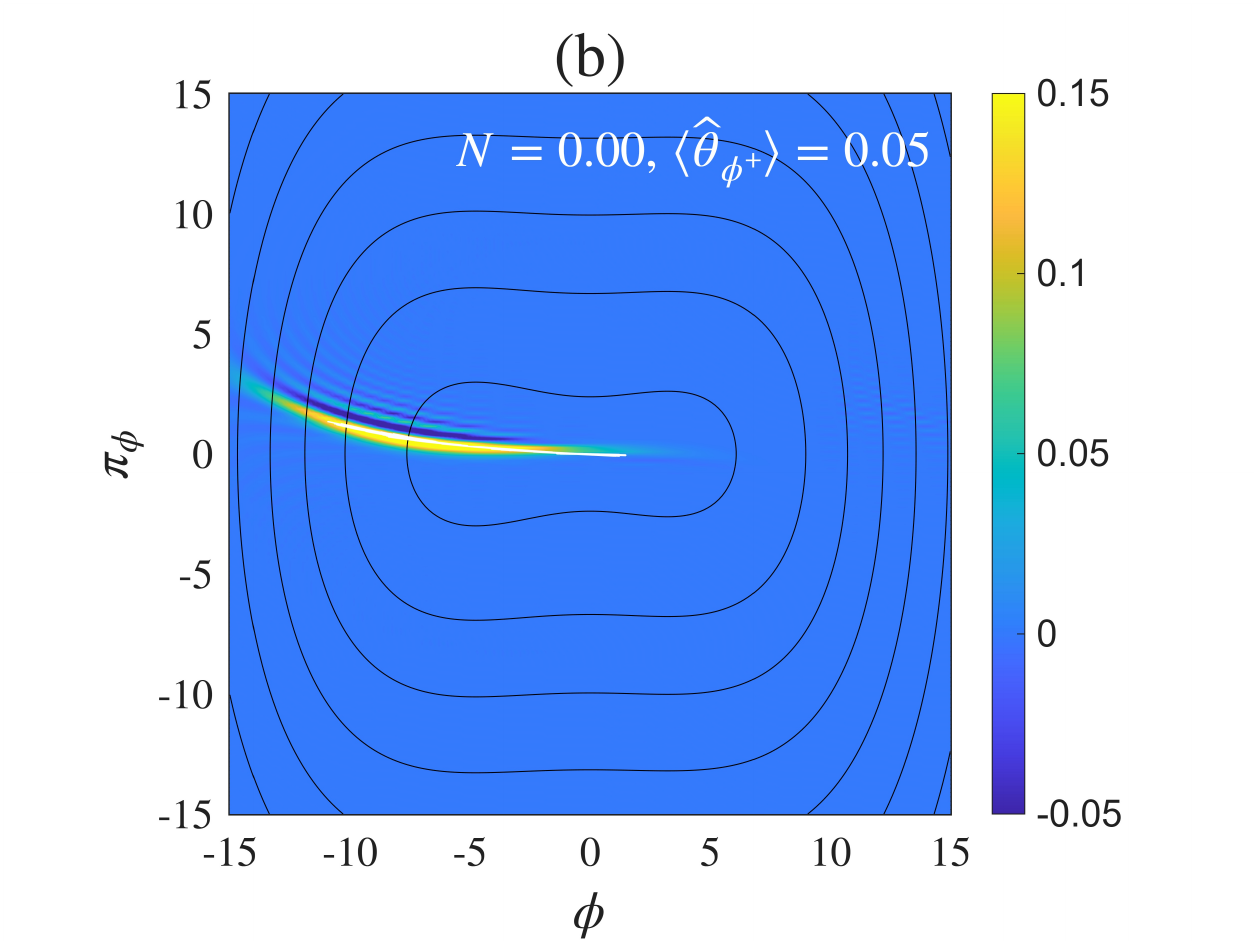}
    \end{subfigure}\hfill
    \begin{subfigure}[t]{0.32\textwidth}\centering
        \panel{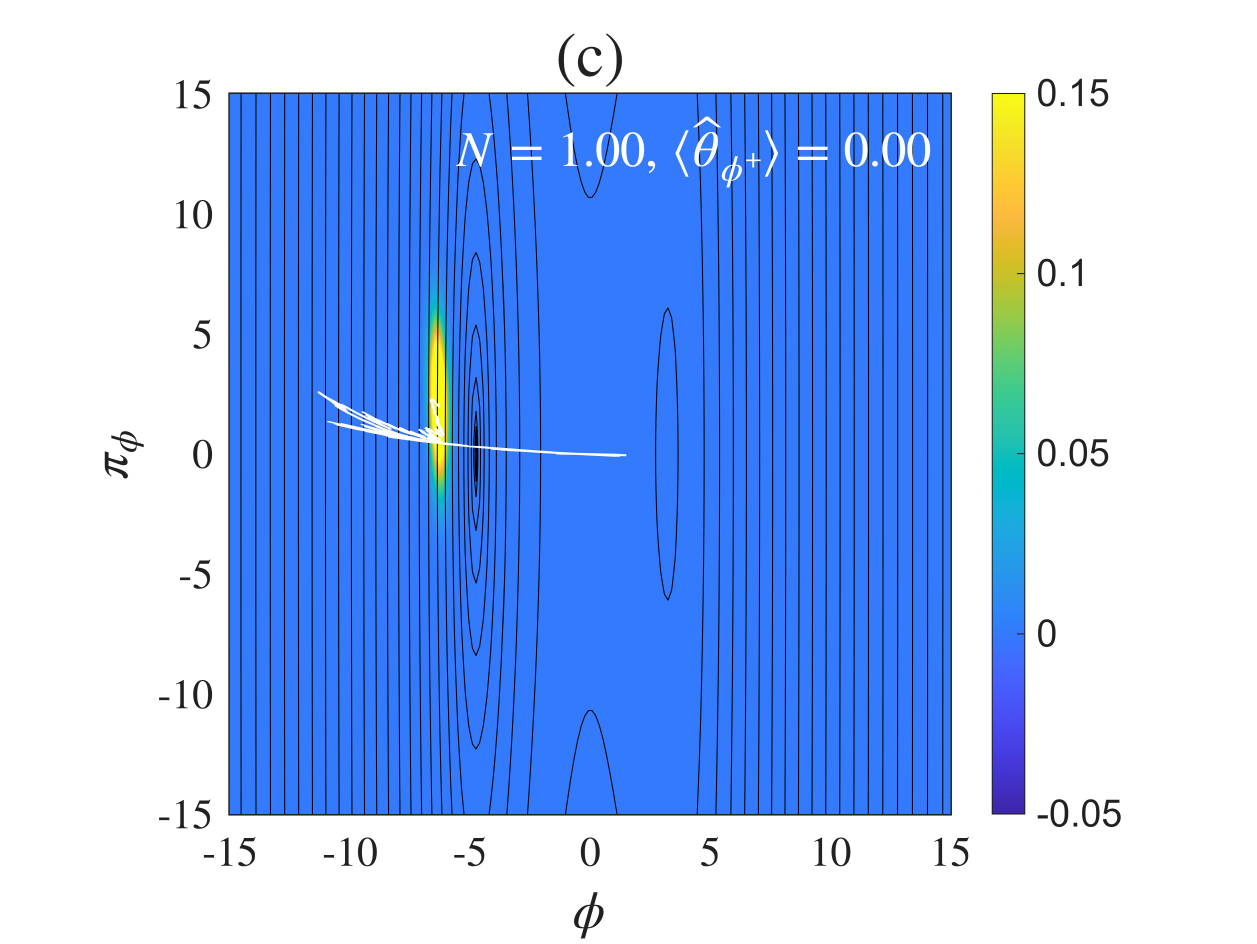}
    \end{subfigure}

    \begin{subfigure}[t]{0.32\textwidth}\centering
        \panel{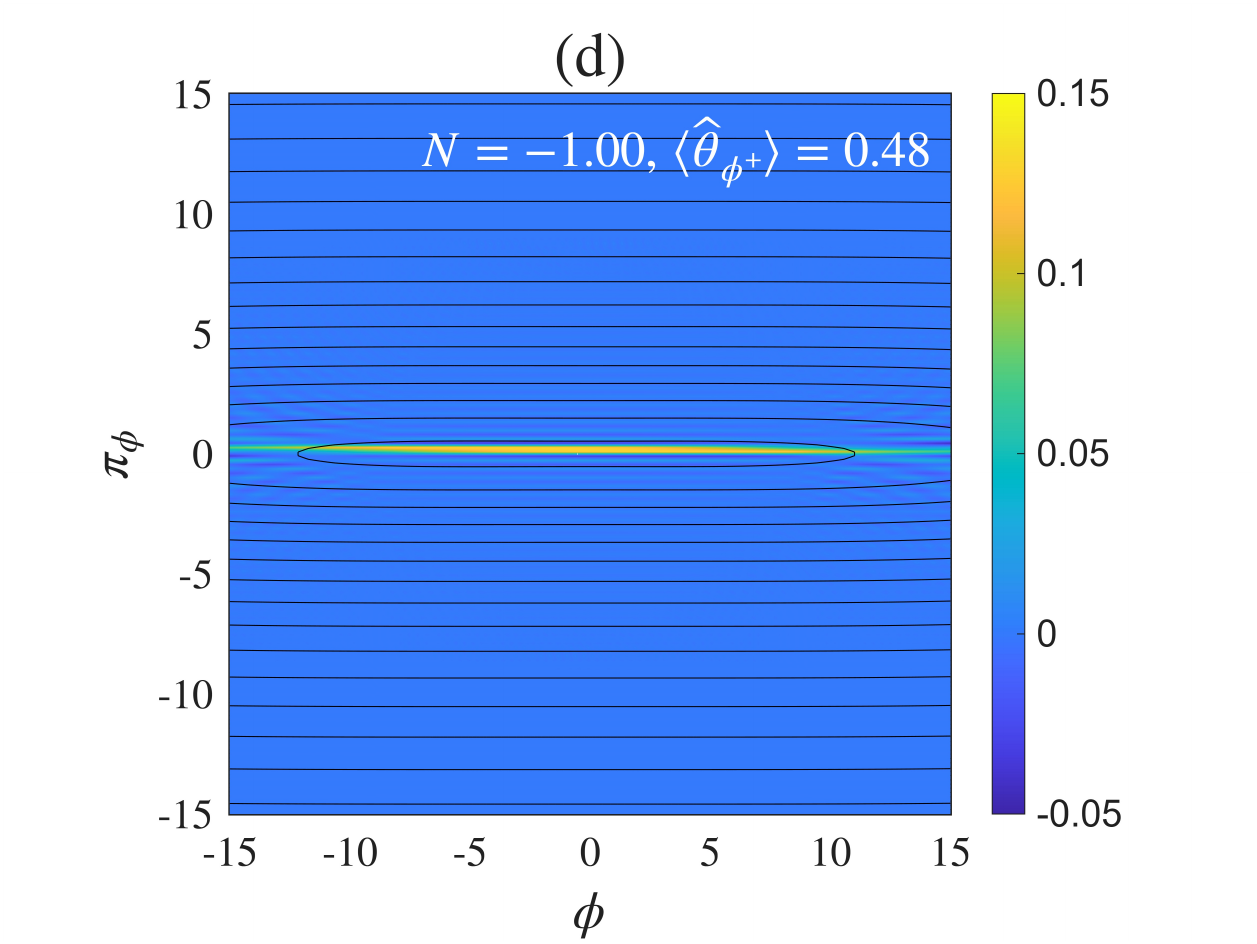}
    \end{subfigure}\hfill
    \begin{subfigure}[t]{0.32\textwidth}\centering
        \panel{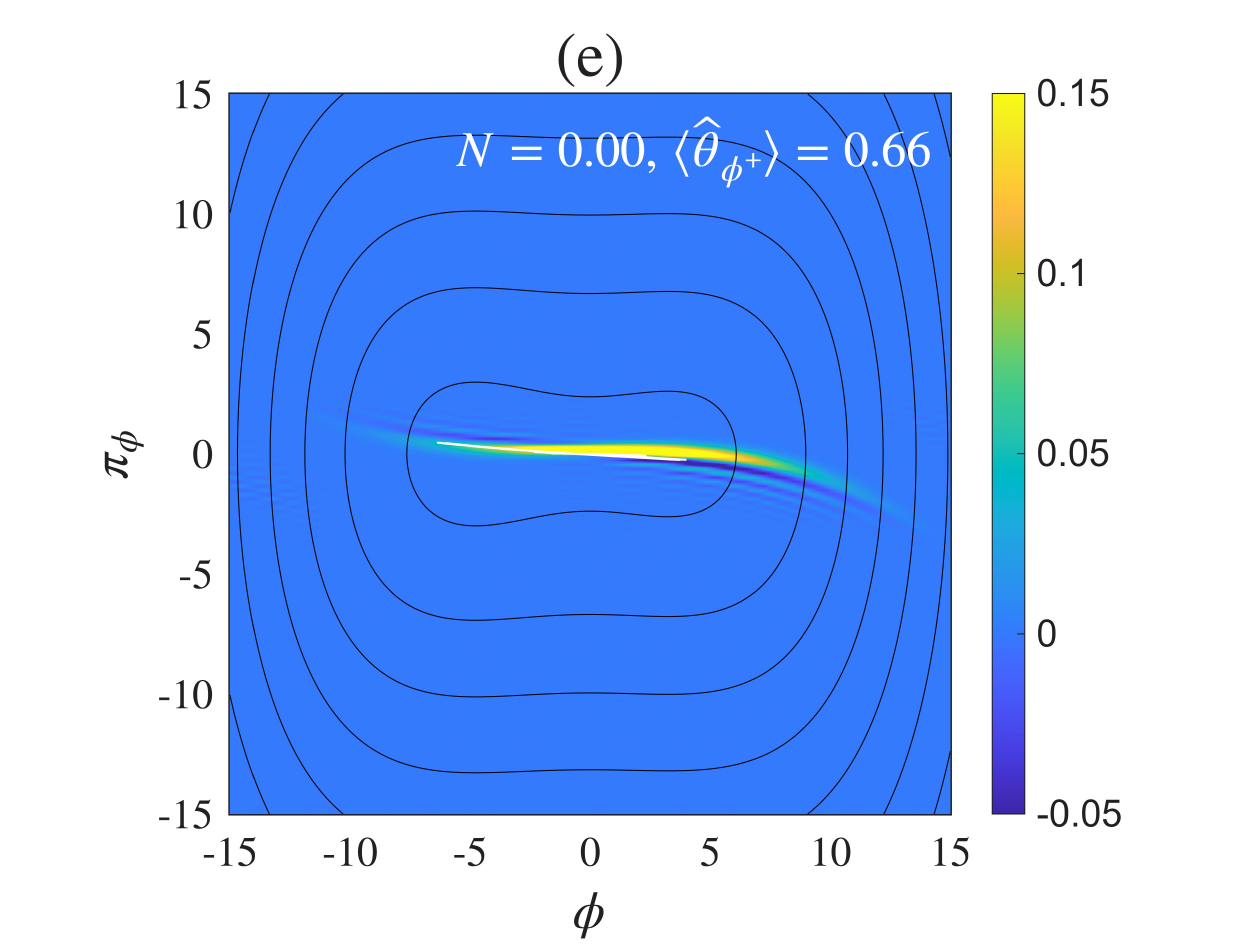}
    \end{subfigure}\hfill
    \begin{subfigure}[t]{0.32\textwidth}\centering
        \panel{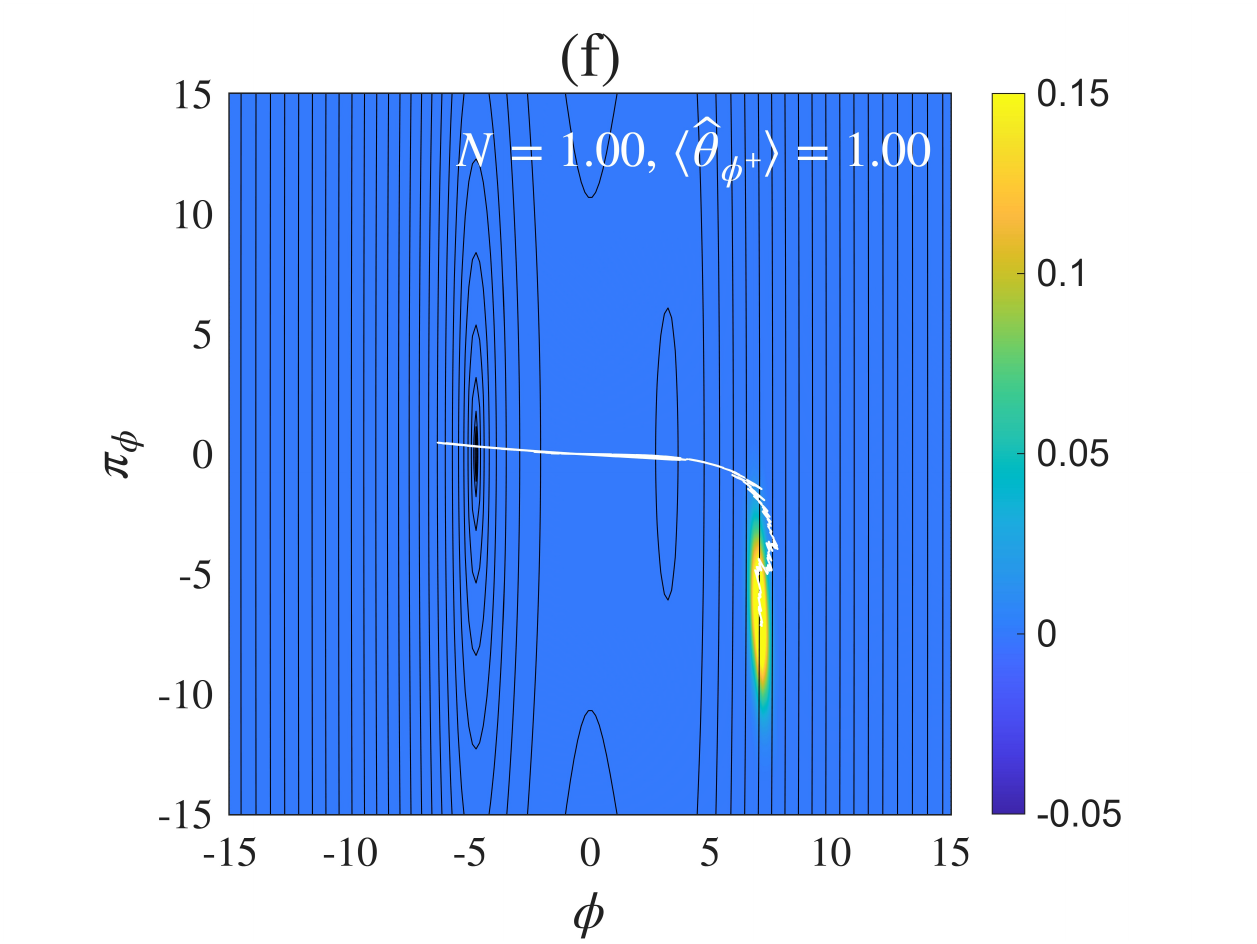}
    \end{subfigure}

    \caption{{\bf SSE dynamics}, \small
    Wigner functions for the SSE evolution~\eqref{eq:SSE_phi_only}.
    The top row shows an SSE trajectory ending in the true vacuum, and the bottom row shows an SSE trajectory ending in the false vacuum. The sonified videos corresponding to these plots are available via these links for the \href{https://youtube.com/shorts/HGWau4xVgx0}{true vacuum} and
    \href{https://youtube.com/shorts/zyUUfPNu2wI}{false vacuum}.}
    .\label{fig:WignerSSE}
\end{figure}

\begin{figure}
    \centering
    \begin{subfigure}[t]{0.32\textwidth}\centering
        \panel{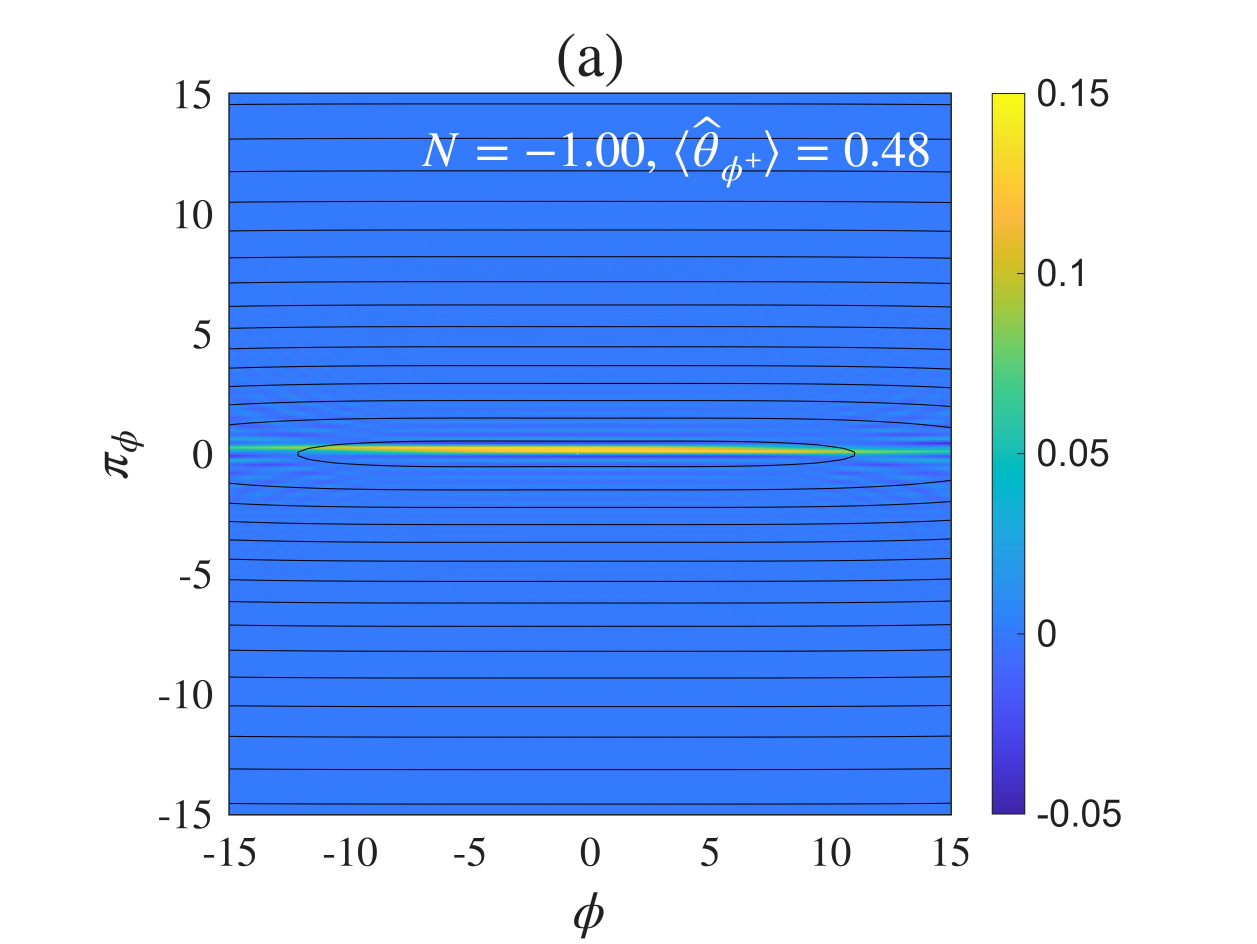}
    \end{subfigure}\hfill
    \begin{subfigure}[t]{0.32\textwidth}\centering
        \panel{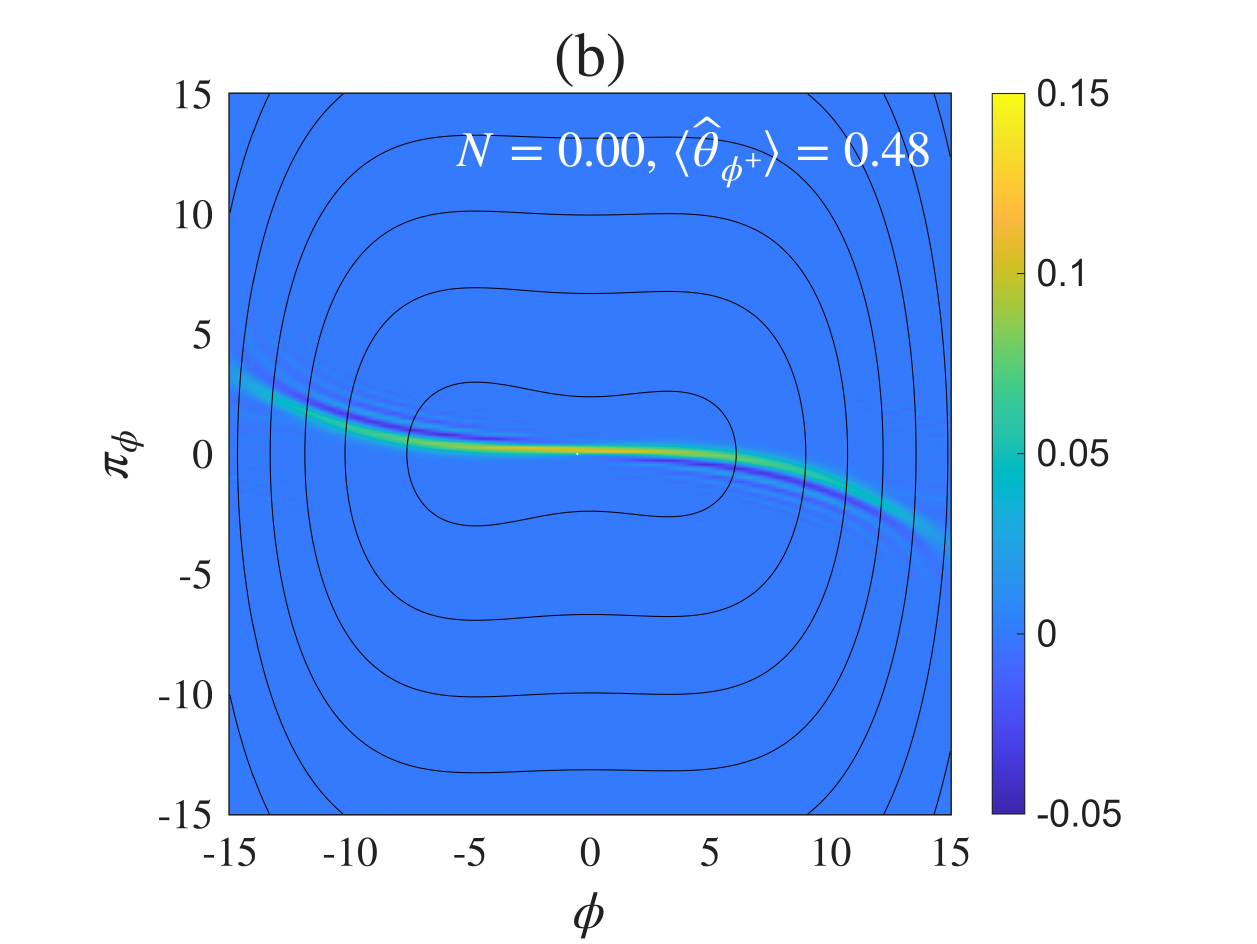}
    \end{subfigure}\hfill
    \begin{subfigure}[t]{0.32\textwidth}\centering
        \panel{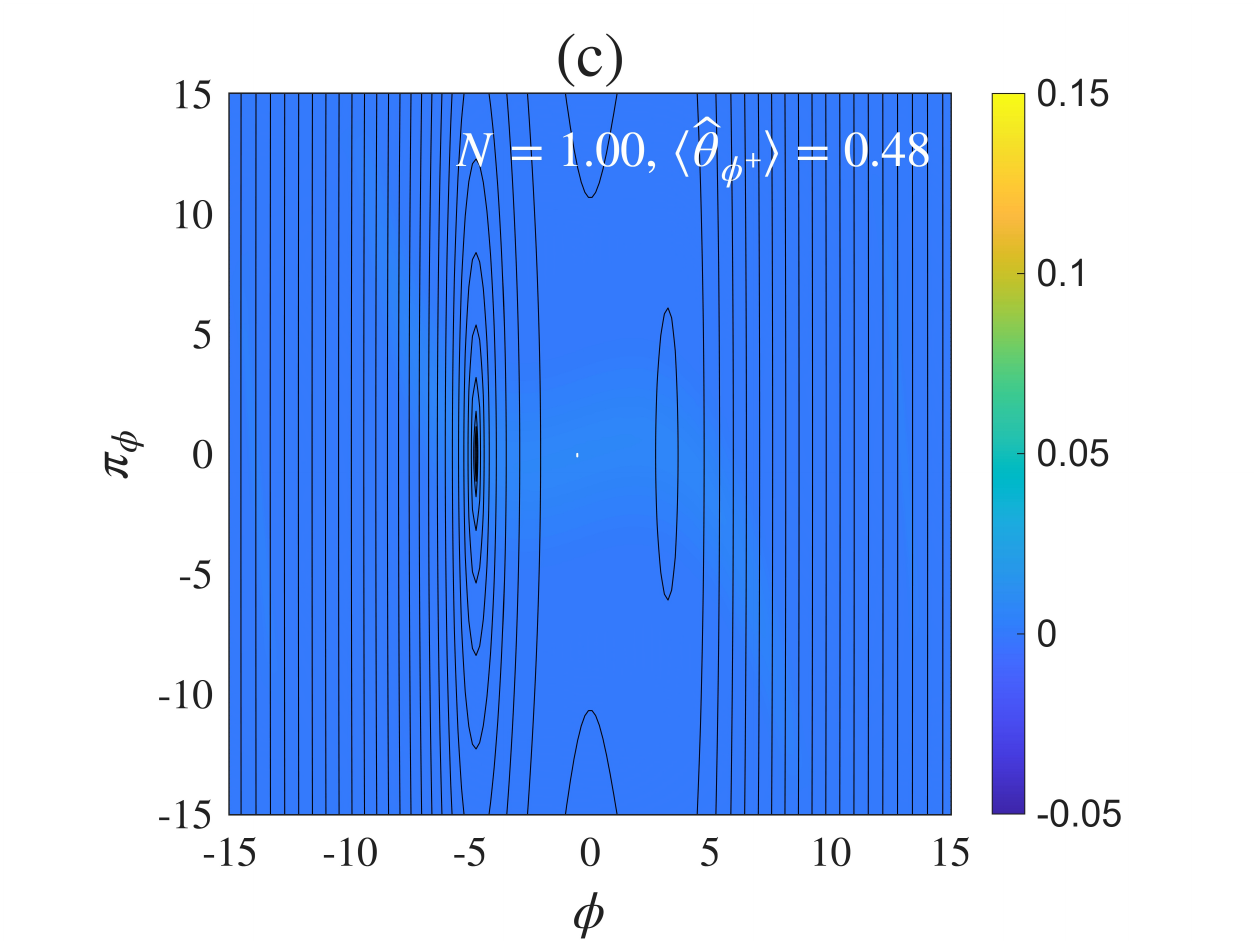}
    \end{subfigure}
    \caption{{\bf GKLS dynamics}, \small
    Wigner functions for Lindblad evolution~\eqref{eq:GKLS_phi_only}. The sonified video corresponding to these plots is available
    \href{https://youtube.com/shorts/Hmx2SgYrrp8}{via this link}.}
    \label{fig:WignerLindblad}
\end{figure}

\begin{figure}[t]
    \centering
    \begin{subfigure}[t]{0.32\textwidth}
        \centering
        \includegraphics[width=\textwidth]{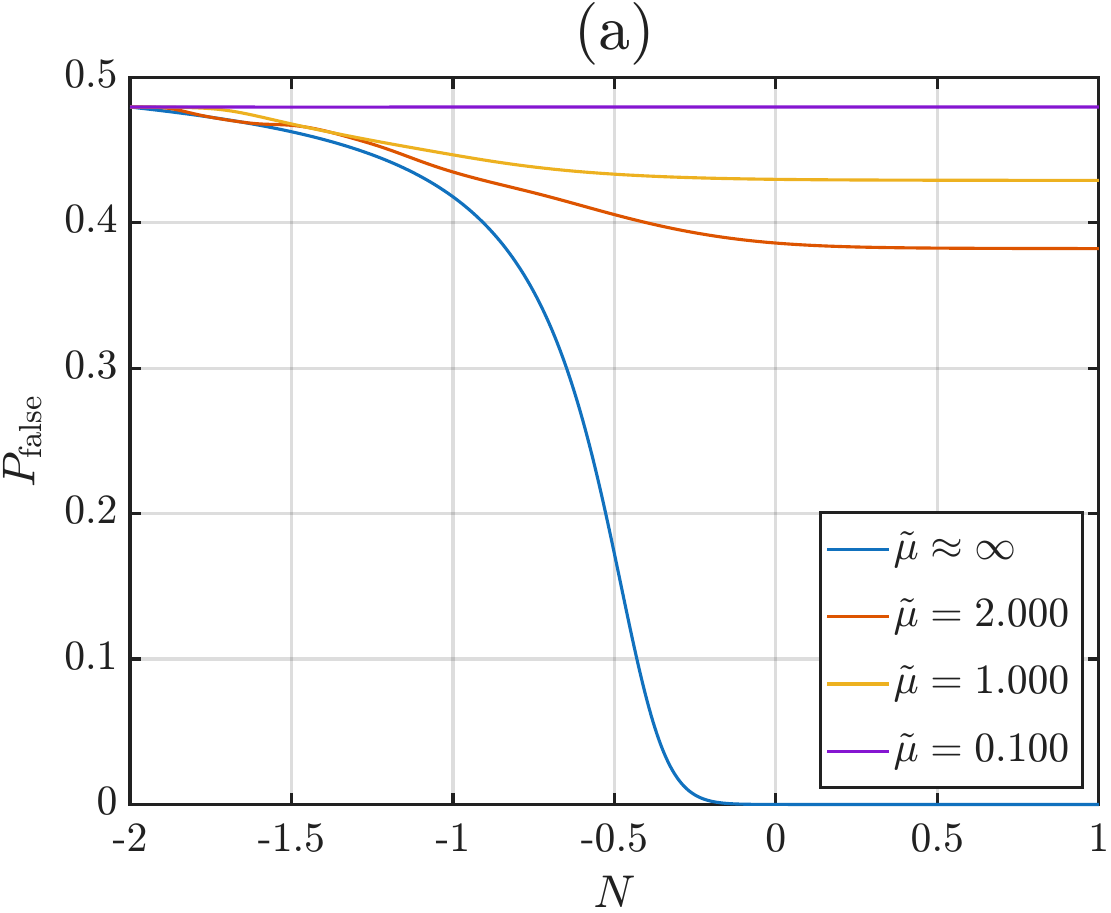}
    \end{subfigure}\hfill
    \begin{subfigure}[t]{0.32\textwidth}
        \centering
        \includegraphics[width=\textwidth]{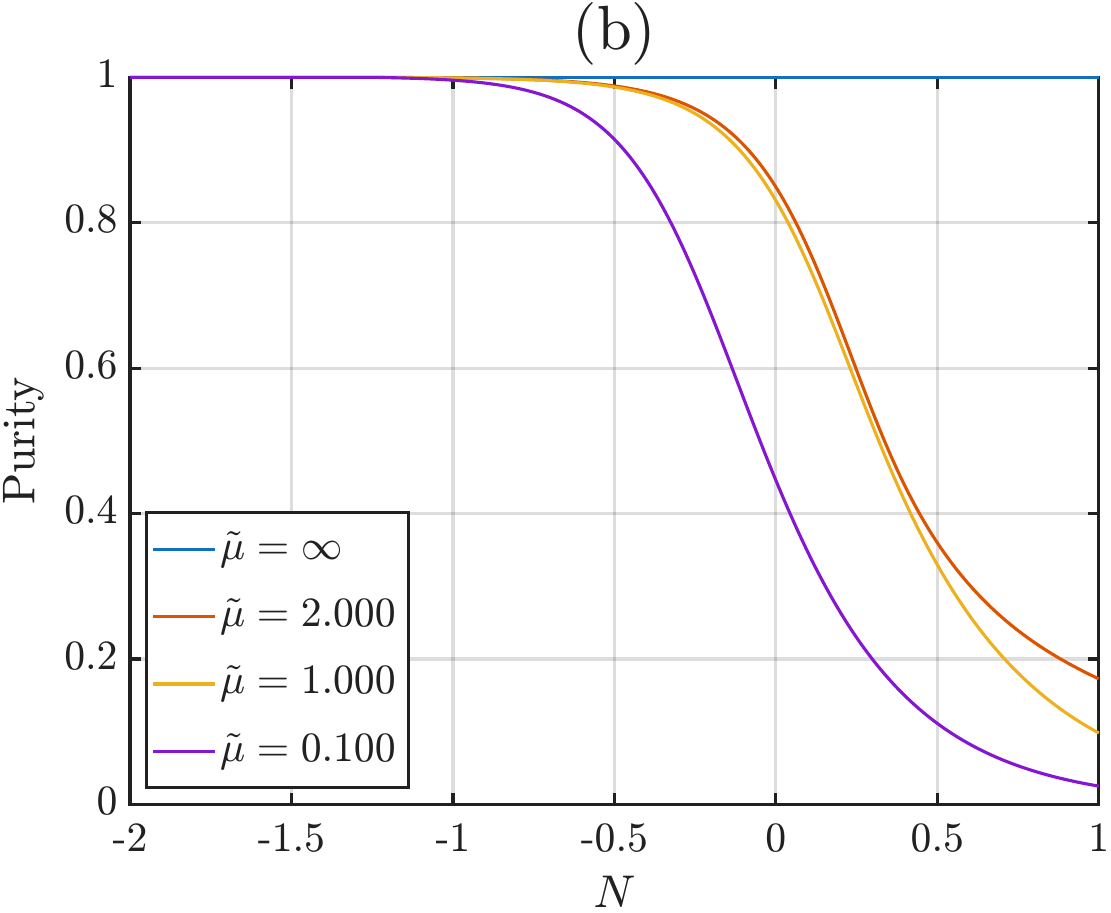}
    \end{subfigure}\hfill
    \begin{subfigure}[t]{0.32\textwidth}
        \centering
        \includegraphics[width=\textwidth]{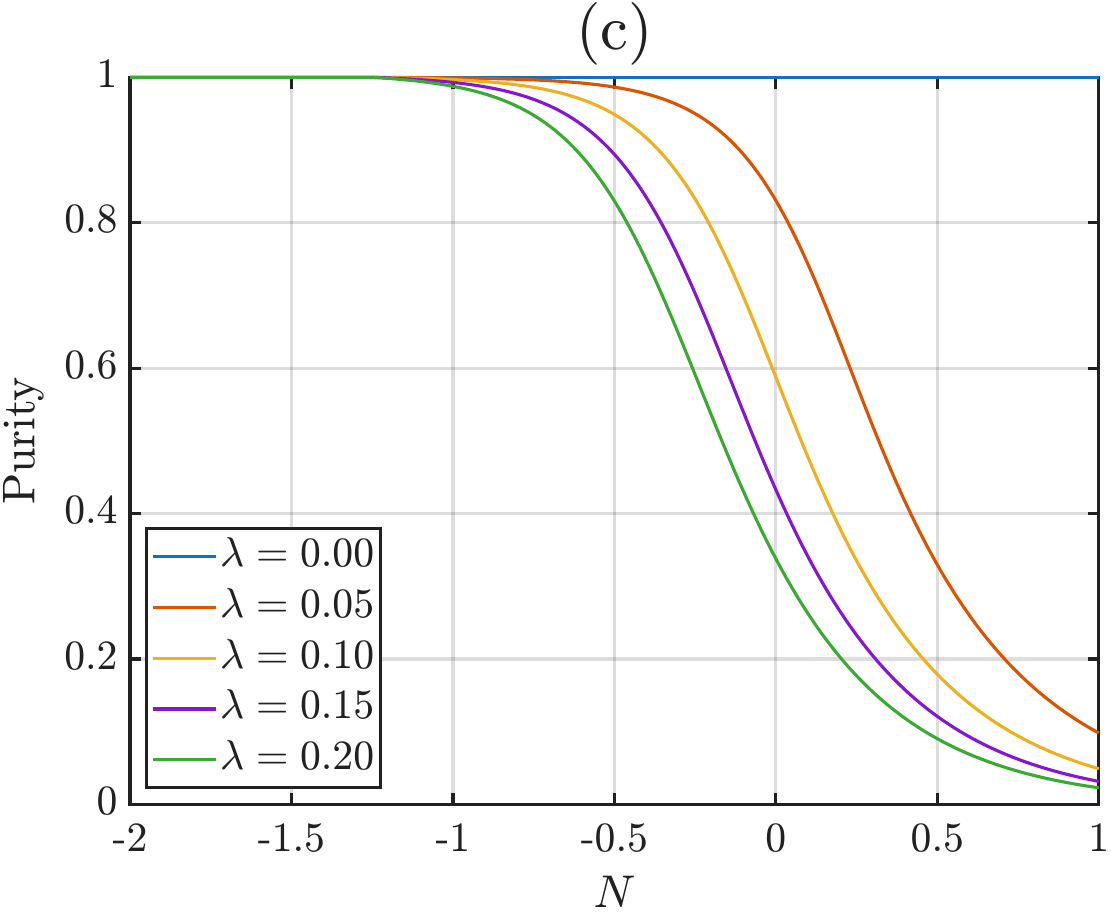}
    \end{subfigure}
    \caption{{\bf  GKLS dynamics}, \small  Parameter-sweep comparison of right-well occupation and purity versus $e$-fold time \(N\) for the GKLS evolution~\eqref{eq:GKLS_phi_only}. (a): adiabaticity \(\tilde{\mu}\) sweep with dephasing strength \(\lambda=0.05\) showing \(P_{\rm false}\equiv \mathrm{Tr} \left(\hat\theta_{\phi^{+}}\hat\rho(N)\right)\). (b): adiabaticity \(\tilde{\mu}\) sweep with \(\lambda=0.05\) showing purity \(P(N)=\mathrm{Tr} \left[\hat\rho^2(N)\right]\). (c): $\lambda$ sweep at fixed adiabaticity \(\tilde{\mu}=1\) showing purity \(P(N)\). We also ran a $\lambda$ sweep for \(P_{\rm false}\) but found no differences in this observable between different $\lambda$ values, as expected. All runs use the same initial state and potential.}
    \label{fig:threefigs}
\end{figure}

We solve the Hilbert-space dynamics introduced in the previous section numerically on the interval \(N\in[-2,1]\), which contains the \emph{barrier switch-on} period during which the effective Hamiltonian transitions from kinetic-term dominance to potential dominance and non-adiabatic effects are most prominent. The initial state used in all simulations is the instantaneous ground state of \(\hat K_S\) at \(N=-2\), shown in Fig.~\ref{fig:WignerAdiabatic}(a). We monitor the right-well occupation
\begin{equation}
\label{eq:Pfalse:def}
P_{\rm false}(N)\equiv \Tr \!\left[\hat\theta_{\phi^{+}}\hat\rho(N)\right],
\qquad
\hat\theta_{\phi^{+}}=\int_{0}^{\infty} \dd \phi \,\ket{\phi}\bra{\phi},
\end{equation}
for which \(P_{\rm false}(N=-2)\simeq 1/2\) due to kinetic dominance at the start. We also track the purity \(\Tr[\hat\rho(N)^{2}]\) and the Wigner function~\cite{wigner1932quantum,wigweyl}
\begin{equation}
	W(\phi,\pi_\phi;N)=\int_{-\infty}^{\infty}d\xi \; \bra{\phi-\tfrac{\xi}{2}}\hat{\rho}(N)\ket{\phi+\tfrac{\xi}{2}}e^{i \pi_{\phi} \xi}.
\end{equation}

Figs.~\ref{fig:WignerAdiabatic}–\ref{fig:WignerLindblad} compare Wigner functions across a range of dynamical scenarios with identical axes, ranges, and colour scales. The black curves show equal-energy contours of the time-dependent Hamiltonian \(\hat K_S(N)\). Columns give snapshots at \(N=-1,0,1\). Fig.~\ref{fig:threefigs} collects false-vacuum occupation and purity for parameter sweeps in the GKLS
dynamics.

At large \(N\), phase-space squeezing requires an exponentially dense grid to resolve the wavefunction accurately, which in our current implementation limits simulations to \(N\le 1\). In the late-time regime (\(N >1\)) the potential term \(e^{3N}V\) dominates over the kinetic term \(e^{-3N}\pi_{\phi}^{2}\), and decoherence is strong (\(\propto e^{6N}\)). In this limit, a strong-localization approximation gives an Arrhenius-type estimate for the stochastic over-the-barrier hopping rates between vacua, which we discuss in the next section and derive in Appendix~\ref{app:GaussSSEArrhenius}.

During the $N\in[-2, 1]$ simulation window, the adiabaticity parameter $\tilde{\mu}$ controls vacuum selection, as demonstrated in Fig.~\ref{fig:threefigs}(a,b). We vary $\tilde{\mu}$ by fixing $\mu = 0.5$ and changing the Hubble rate, which keeps the initial state identical between runs; changing $\mu$ at fixed $H$ would instead modify the ground state at $N = -2$ and complicate a direct comparison. For large $\tilde{\mu}$ the evolution is nearly adiabatic: the state closely follows the instantaneous ground state into the deeper (true) vacuum, giving $P_{\rm false} \simeq 0$. For small $\tilde{\mu}$ the evolution becomes strongly non-adiabatic: the barrier \emph{switches on} faster than the state can adjust, freezing population in an excited configuration so that the true vacuum is only weakly favored. As shown in Fig.~\ref{fig:threefigs}(a), decreasing $\tilde{\mu}$ therefore increases $P_{\rm false}$ at late times for both Schr\"odinger and GKLS dynamics.

Fig.~\ref{fig:threefigs}(b) shows that in the adiabatic limit the state remains a pure ground state, with the rate of purity loss increasing as $\tilde{\mu}$ decreases. This is due to the larger Hubble rates in these runs non-adiabatically exciting higher states with subsequent decoherence diagonalizing the density operator. Turning on decoherence by increasing $\lambda$ from $0$ to $0.2$ leaves $P_{\rm false}(N)$ essentially unchanged but markedly increases the rate of purity loss; see Fig.~\ref{fig:threefigs}(c).

\begin{description}
    \item[\textbf{Instantaneous ground state, Fig.~\ref{fig:WignerAdiabatic}:}] 
    At \(N=-2\) the state is a stretched Gaussian, effectively delocalized in \(\phi\) when measured against the size of the barrier. By \(N=0\) the instantaneous ground state is approximately Gaussian and circular with comparable variances in \(\phi\) and \(\pi_\phi\) located in the basin of the true vacuum with $P_{\rm false}\simeq 0$. By \(N=1\) the state is strongly squeezed and localized at the true minimum.

    \item[\textbf{Schr\"odinger equation, Fig.~\ref{fig:WignerHam}:}] 
    The delocalized Wigner function non-adiabatically shears and develops interference fringes between the true and false vacua, while the purity remains \(1\). The state eventually freezes into a superposition over both vacua, with a false-vacuum occupation probability \(P_{\rm false} \simeq 0.48\).

    \item[ \textbf{SSE trajectories, Fig.~\ref{fig:WignerSSE}:}] 
    The bath-induced dephasing channel acts as a continuous measurement of \(\phi\). Each pure trajectory rapidly localizes into one well; fringes are suppressed and the Wigner function collapses to a highly squeezed and approximately Gaussian wavepacket. \(P_{\rm false}\) drifts toward \(0\) or \(1\) depending on the selected well. The momentum expectation value can take large excursions due to a random walk in an almost flat \(\pi_{\phi}\) direction.

    \item[ \textbf{GKLS dynamics, Fig.~\ref{fig:WignerLindblad}:}] 
    The GKLS evolution produces a marginal in \(\phi\) probability density that is very similar to that of the Schr\"odinger dynamics. However, bath-induced decoherence strongly suppresses nonclassical correlations (fringing and regions of Wigner-function negativity~\cite{hudson1974wigner,Wignegativity}). By the end of the simulation interval, the Wigner function describes a highly mixed state with essentially the same false-vacuum occupation probability \(P_{\rm false} \simeq 0.48\) as the closed-system dynamics.

    \item[\textbf{Parameter sweeps, Fig.~\ref{fig:threefigs}:}]
    Panel~(a) shows that decreasing the adiabaticity parameter \(\tilde\mu\) (at fixed initial state) increases late-time \(P_{\rm false}\) for GKLS dynamics \eqref{eq:GKLS_phi_only}, reflecting non-adiabatic \emph{barrier switch-on} that freezes population in an excited configuration. Panel~(b) shows that purity loss accelerates as \(\tilde\mu\) decreases, as non-adiabaticity generates superpositions across both vacua, which are decohered more rapidly than localized states. We found that increasing \(\lambda\) at fixed \(\tilde\mu\) leaves \(P_{\rm false}(N)\) essentially unchanged yet significantly speeds up purity loss as evidenced in Panel~(c).
\end{description}

\noindent In the next section we show that after the non-adiabatic switch-on phase, the subsequent expansion drives a ‘cosmic lockdown’ in which the false/true occupation becomes dynamically stable.

\section{Late-time cosmic lockdown}
In the previous section we showed that non-adiabatic dynamics during the kinetic-to-potential transition can substantially enhance the false-vacuum occupation of a coarse-grained light spectator field. The natural question is what happens subsequently, over the many e-folds that may remain until the end of inflation. In this section we argue that once the state has localized into a definite well during the barrier switch-on, the bath-induced monitoring rapidly suppresses inter-well coherences and thereby strongly inhibits further quantum tunneling. Any remaining relaxation of the false-vacuum population must then proceed through rare stochastic over-the-barrier hops, whose rate is
exponentially suppressed in the late-time, strong-decoherence regime. In this
sense the system becomes effectively \emph{locked} into the stochastically
selected minimum.

As discussed in \S\ref{subsec:fourpoint_Lindblad_SSE}, the coarse-grained
spectator dynamics is governed by the normalised stochastic Schr\"odinger
equation
\begin{multline}
\dd\ket{\psi}
= - i\left[\frac{e^{-3N}}{2H\mathtt{vol}}\,\hat \pi_\phi^{2}
           +\frac{e^{3N} \mathtt{vol}}{H}\,V(\hat \phi) \right]\ket{\psi}\, \dd N
  \\-\frac{131 \pi\lambda^{2} e^{6N}}{512 \mu^5 \mathtt{vol}}
   \big(  \hat \phi-\langle \hat \phi\rangle \big)^2\ket{\psi}\,\dd N
+\sqrt{\frac{131 \pi\lambda^{2} e^{6N}}{256 \mu^5 \mathtt{vol}}}
    \big(\hat \phi-\langle \hat \phi \rangle\big)\ket{\psi}\; \dd W.
\label{eq:appSSE_norm_bathonly}
\end{multline}
The localizing term grows as
$e^{6N}$, so at sufficiently large $N$ the dynamics is dominated by rapid
dephasing in the $\phi$ pointer basis. The state is driven into a narrow
wavepacket in field space, and sustained inter-well tunneling would require
the continual regeneration of coherences that the Lindblad channel
immediately destroys. The late-time evolution is therefore well described by a
classical stochastic process for the localized wavepacket center, with
exponentially rare barrier-crossing events.

As detailed in Appendix~\ref{app:GaussSSEArrhenius}, in the strong-localization regime the stochastic dynamics may be truncated at Gaussian order by evolving the first moments together with the covariance matrix and then reducing to an effective overdamped Langevin equation for the wavepacket center. This leads to an Arrhenius/Eyring--Kramers estimate for the mean first-passage time (measured in $e$-folds from the crossover time Eq~\eqref{eq:crossover}) from the false to the true vacuum,
\begin{equation}
\left\langle N_{F \rightarrow T}\right\rangle \simeq
\frac{6 \pi H^2}{\mu^2}\,
\sqrt{\frac{\beta_4+\beta_3}{2 \beta_4}}\,
\exp \left[
\frac{128}{131 \pi}\,
\frac{3 \beta_4+\beta_3}{\left(\beta_4+\beta_3\right)^3}\,
\frac{\mathtt{vol}^3 \mu^9}{\lambda^2}
\right].
\label{eq:NFtoT_meanFPT}
\end{equation}
The exponential can be expressed in terms of the dimensionless coarse-graining measure $\mu^{3}\mathtt{vol}$, which corresponds to the physical coarse-graining volume evaluated at crossover in units of the microscopic length scale $\mu^{-1}$. The mean first-passage time is therefore exponentially sensitive to $\left(\mu^{3}\mathtt{vol}\right)^{3}$. Consequently, once decoherence has localized the state within a single well, subsequent inter-well relaxation of the non-adiabatically enhanced false-vacuum population is generically inefficient unless the system-bath coupling is very large or the effective coarse-graining volume is very small.

\FloatBarrier

\section{Summary}

We have derived a coarse-grained description of spectator scalar fields defined on a finite comoving patch, and used it to study the interplay between non-adiabatic dynamics and decoherence in an asymmetric double-well potential. Although simplified, this setup illustrates generic mechanisms that are expected to operate in more realistic cosmological settings and are often neglected. When the field is light compared to the Hubble scale, the background expansion occurs too rapidly for the field to adiabatically follow, driving excitations out of the instantaneous ground state and producing a superposition delocalized across both vacua. Interactions with additional spectator fields then rapidly decohere this superposition, selecting a definite vacuum in field space.

After decoherence, inter-well tunneling is strongly inhibited. Within our finite simulation window, the false-vacuum occupation saturates under GKLS evolution. Over-the-barrier hops from the false to the true vacuum can, in principle, relax this excitation, but they are exponentially rare; as a result, inflation may end with an enhanced false-vacuum occupation.

\section{Data and code availability}
Matlab and Mathematica codes to generate all figures and data in the paper can be found on GitHub: \url{https://github.com/rchristie95/CosmicLockdownMatlabAndMathematica}. Sonified videos of the Wigner function dynamics can be found on YouTube: \url{https://www.youtube.com/playlist?list=PLnFRudoWkGcH35xIeObYtbGmptvwb7CH9} and sonification method is given in~\cite{christie2024sound}.

\section{Acknowledgements}

We thank Jason Pollack, Sarah Shandera and Varun Vaidya for helpful discussions.
This work was supported by the Science and Technology Facilities Council (grant number ST/W001225/1).
For the purpose of open access, the authors have applied a Creative Commons Attribution (CC-BY) licence to any Author Accepted Manuscript version arising from this work.
Supporting research data are available on reasonable request from the corresponding author, Greg Kaplanek.

\bibliographystyle{JHEP}
\bibliography{main.bib}

\newpage
\appendix
\section{Master equations and bath correlators in de Sitter}
\label{app:derivation}

This appendix derives the time-local Gorini-Kossakowski-Lindblad-Sudarshan (GKLS) equation governing the infrared dynamics of a light scalar test field in de Sitter space. Throughout, we restrict attention to the \emph{coarse-grained fields} in a fixed comoving box.

\subsection{Single-oscillator bath in de Sitter}
\label{app:truncation}

We begin with the four-dimensional action
\begin{equation}
  S = - \int \mathrm{d}^{4}x \, \sqrt{-g} \left(\mathcal{L}_{\phi}+\mathcal{L}_{\chi}-\mathcal{L}_{\rm int}^{(k)}\right)
  \label{eq:A_action}
\end{equation}
with the system and environment Lagrangian densities
\begin{equation}
  \mathcal{L}_{\phi}=\frac{1}{2}(\partial \phi)^2-V(\phi) \qquad\text{and}\qquad \mathcal{L}_{\chi}=\frac{1}{2}(\partial \chi)^2-\frac{m^2}{2} \chi^2\, ,
\end{equation}
and system potential given in \eq{eq:potential}. We consider various four point interactions 
\begin{equation} \label{eq:quartic_interactions}
    \mathcal{L}_{\rm int}^{(1)}=\underline{\lambda}_{1} \phi^{3} \chi,\qquad \mathcal{L}_{\rm int}^{(2)}=\underline{\lambda}_{2} \phi^{2} \chi^{2},\qquad \mathcal{L}_{\rm int}^{(3)}=\underline{\lambda}_{3} \phi \chi^{3}
\end{equation}
where $\underline{\lambda}_{i}$ are bare (denoted by the under bar) dimensionless coupling strengths. To isolate the long-wavelength (infrared) dynamics relevant for stochastic and open-system treatments, we coarse-grain the system over a fixed comoving patch $R$ with volume $\mathtt{vol}$. Assuming the fields are approximately homogeneous within the box we may neglect gradient terms
\begin{equation}
  \phi(x,N) \;\to\; \phi(N) \equiv \frac{1}{\mathtt{vol}}\int_R \mathrm{d}^3x \,\Phi(\mathbf{x},N),
  \qquad
  \chi(x,N) \;\to\; \chi(N) \equiv \frac{1}{\mathtt{vol}}\int_R \mathrm{d}^3x \,X(\mathbf{x},N).
\end{equation}
This yields a Lagrangian describing two coupled oscillators evolving in de Sitter ~\cite{misner1973minisuperspace,kiefer2022quantum},
\begin{equation}
  L(t)=\mathtt{vol} \, a^{3}(t)\left[
           \frac{\dot{\phi}^{2}}{2}-V(\phi)
          +\frac{\dot{\chi}^{2}}{2}-\frac{m^{2}}{2}\chi^{2}
          -\mathcal{L}_{\rm int}^{(k)}
        \right].
  \label{eq:A_L_zero}
\end{equation}
To pass to Hamiltonian language we define canonical momenta in the
usual way,
\begin{equation}
  \pi_\phi\equiv\frac{\partial L}{\partial\dot\phi}=\mathtt{vol}\,a^{3}\dot\phi,
  \qquad
  \pi_\chi\equiv\frac{\partial L}{\partial\dot\chi}=\mathtt{vol}\,a^{3}\dot\chi,
  \label{eq:A_momenta_t}
\end{equation}
and after a brief calculation obtain the Hamiltonian function
\begin{equation}
  \mathcal{H}=\mathtt{vol}\,a^{3}(t)\left[
           \frac{\dot{\phi}^{2}}{2}+V(\phi)
          +\frac{\dot{\chi}^{2}}{2}+\frac{m^{2}}{2}\chi^{2}
          +\mathcal{L}_{\rm int}^{(k)}
        \right].
  \label{eq:A_H_zero}
\end{equation}
For simulations $N$ is more convenient than cosmic time $t$,
\(a(t)=e^{Ht}=e^{N}\) and therefore \(\dot{\phi}=H {\phi}'\).  In terms of
$N$ the canonical momenta become
\begin{equation}
  \pi_\phi=e^{3N}H\mathtt{vol} \phi',
  \qquad
  \pi_\chi=e^{3N}H\mathtt{vol} \chi',
  \label{eq:A_momenta_N}
\end{equation}
and the box-averaged Hamiltonian operator reads
\begin{equation}
   \hat K(N)=\frac{e^{-3N}}{2H \mathtt{vol}}\hat \pi_\phi^{2}
           +\frac{e^{3N}\mathtt{vol}}{H}V(\hat \phi)
           +\frac{e^{-3N}}{2H\mathtt{vol}} \hat \pi_\chi^{2}
           +\frac{m^{2}_{\chi}\mathtt{vol} e^{3N}}{2H}\hat \chi^{2}
           +\frac{\mathtt{vol} e^{3N}}{H}\mathcal{L}_{\rm int}^{(k)} .
  \label{eq:A_K}
\end{equation}
We can diagonalise the bath Hamiltonian 
which leads to the compact expression for the bath $\chi$ oscillator Hamiltonian
\begin{equation}
  \hat K_{\rm B}^{(\chi)}(N)
  =\omega_\chi\left[\hat b^{\dagger}(N)\hat b(N)+\frac12\right]
   \quad\text{with}\quad
  \omega_\chi\equiv\frac{m_{\chi}}{H} \quad\text{and}\quad \hat b(N)=\sqrt{\frac{m_{\chi} \mathtt{vol}e^{3N}}{2}}\chi
        +\frac{i}{\sqrt{2m_{\chi} \mathtt{vol} e^{3N}}}\hat \pi_\chi.
  \label{eq:A_ladder}
\end{equation}
If we assume that the system and bath are coupled at $N=N_0$ we can write down a Nakajima-Zwanzig equation \cite{petruccione} in the interaction picture (operators acquire a
subscript \(I\)) with the reduced density matrix of the $\phi$ field obeying
\begin{equation}
  \dot\rho_{I}(N)
   =-  \underline{\lambda}_k^2\int_{N_{0}}^{N}  \dd \bar N\;
      C_{k}(N,\bar N) 
      [\hat S_{I}^{(k)}(N),[\hat S_{I}^{(k)}(\bar N),\rho_{I}]]\, .
  \label{eq:A_exact_ME}
\end{equation}
We have split the interaction term $\hat{S}_I^{(k)} \otimes \hat{B}_I^{(k)}$ as 
\begin{equation}
  \hat{S}_I^{(k)}(N) = \frac{\mathtt{vol} e^{3N}}{H} \, \hat{\phi}_I^{4-k}(N), 
  \quad \text{and} \quad 
  \hat{B}_I^{(k)}(N) = \underline{\lambda}_{k} \, \hat{\chi}_I^{k}(N)\label{eq:A_L_int}
\end{equation}
and the bath enters only through the fluctuation correlation functions: 
\begin{equation}
  C_{k}(N,\bar N)=
   \bigl\langle\mathbf{0}_{N_0}| \hat{\chi}_I^{k}(N)\hat{\chi}_I^{k}(\bar N)|\mathbf{0}_{N_0}\bigr\rangle-\bigl\langle\mathbf{0}_{N_0}|\hat{\chi}_I^{k}(N)|\mathbf{0}_{N_0}\bigr\rangle\bigl\langle\mathbf{0}_{N_0}| \hat{\chi}_I^{k}(\bar N)|\mathbf{0}_{N_0}\bigr\rangle.
\end{equation}
As the $\chi$ field evolves under a quadratic Hamiltonian we may use Isserlis' theorem \cite{pavliotis2014stochastic} to write the higher fluctuation correlation functions in terms of the two-point function $C_1$ as
\begin{equation}
  C_{2}(N,\bar N)=2  C_1(N,\bar N)^2\quad\text{and}\quad C_{3}(N,\bar N)=9 C_1(N,N) C_1(\bar N, \bar{N}) C_1(N,\bar N)+6 C_1(N, \bar N)^3 \, .\label{eq:isserlis}
\end{equation}
To evaluate this two-point function we start by expressing the interaction picture $\chi$ operator in terms of ladder operators as
\begin{equation}
\hat{\chi}_{I}(N) =
\frac{
  e^{-\frac{3}{2} N}
}{
  \sqrt{2 \omega_{\chi} H \mathtt{vol} }}\left[
  \hat b(N)
  e^{- i \omega_{\chi} \Delta}
  +
  \hat b^{\dagger}(N)e^{ i \omega_{\chi} \Delta}
\right]
\end{equation}
where we introduce the time differences \(\Delta=N-N_{0}\) and \(\bar\Delta=\bar N-N_{0}\) for brevity. We can write the interaction picture field operator in terms of ladder operators at $N_0$ as
\begin{multline}
    \hat{\chi}_{I}(N) =
\frac{
  e^{-\frac{3}{2} N}
}{
  \sqrt{2  \omega_{\chi}  H  \mathtt{vol}}}\Bigg\lbrace
  \left[
    \cosh\left( \frac{3 \Delta}{2} \right) \hat{b}(N_0)
    + \sinh\left( \frac{3 \Delta}{2} \right) \hat{b}^\dagger(N_0)
  \right]
  e^{- i \omega_{\chi} \Delta}
  \\+
  \left[
    \cosh\left( \frac{3 \Delta}{2} \right) \hat{b}^\dagger(N_0)
    + \sinh\left( \frac{3 \Delta}{2} \right) \hat{b}(N_0)
  \right]
  e^{ i \omega_{\chi} \Delta}
\Bigg\rbrace
\end{multline}
by making use of the commutation relations
\begin{equation}
  [b(N),b^{\dagger}(N_0)]=\cosh\Bigl(\frac32\Delta\Bigr),
  \qquad
  [b(N),b( N_0)]=-\sinh\Bigl(\frac32\Delta\Bigr),  \qquad
[b^{\dagger}(N),b^{\dagger}( N_0)]=\sinh\Bigl(\frac32\Delta\Bigr).
\end{equation}
The two point function is thus given by
\begin{multline}
  C_1(N,\bar N) =\frac{e^{-\frac{3}{2} (N + \bar{N})}}{
2 H \omega_{\chi} \mathtt{vol}
}
\left\lbrace
\cosh\!\left[\frac{1}{2}\Delta (3 + 2 i \omega_{\chi})\right]
+ \sinh\!\left[\frac{1}{2} \Delta(3 - 2 i \omega_{\chi})\right]
\right\rbrace\\
\left\lbrace
\cosh\!\left[\frac{1}{2} \bar{\Delta}(3 - 2 i \omega_{\chi})\right]
+ \sinh\!\left[\frac{1}{2} \bar{\Delta}(3 + 2 i \omega_{\chi})\right]
\right\rbrace
  \label{eq:A_C_exact}
\end{multline}
with higher correlators $C_2$ and $C_3$ given by \eq{eq:isserlis}.

\subsection{Continuous oscillator bath spectra} 
To include a continuum of heavy modes we replace the single
oscillator \(\chi\) with a set \(\{\chi_{m}\}\) labelled by their
mass $m$.  The bulk Lagrangian is therefore generalised to  
\begin{equation}
  L(t) =
  \mathtt{vol} \cdot a^{3} \Biggl[
      \frac{ \dot{\phi}^{2} }{2}-V(\phi)
     - \int_{0}^{\infty} \frac{\dd m}{\mu}\,
       \Bigl( \frac{\dot{\chi}_m^{2}}{2}  
          -\frac12 m^{2}\chi_m^{2} + \underline{\lambda}_k g^{k/2}(m)  \,\phi^{4-k}\chi^k_m        
       \Bigr)
     \Biggr] \quad\text{with}\ \ 
g(m) = \frac{m}{\mu} e^{-\tfrac{m}{\Lambda}}.
  \label{eq:L_ohmicApp}
\end{equation}
The form of the spectral density $g(m)$ is introduced to compensate the mass scaling of the correlators, \(C_k(N,\bar N;m)\propto m^{-k}\), so that the integrand in \eqref{eq:L_ohmicApp} results in Markovian white noise rather than power–law suppressed coloured noise. $\Lambda$ is a high mass cutoff which acts as a regulator as we later take $\Lambda\to \infty$. In the remainder of this appendix we derive renormalised Markovian master equations corresponding to the three $\mathcal{L}$ choices.

\subsection*{$\phi^3 \chi$ interaction:}
The continuous two point correlation function is thus
\begin{equation}
  C_1^{\rm cont}(N,\bar N,\Lambda)
  = \frac{1}{\mu^{2}}
    \int_{0}^{\infty} \dd m \int_{0}^{\infty} \dd m'\,
      \sqrt{g(m) g(m')}
      \langle \mathbf{0}_{N_0}|
        \chi_{m,I}(N)\,\chi_{m',I}(\bar N)
      |\mathbf{0}_{N_0}\rangle .
\end{equation}
Environmental fields are uncorrelated in mass
\(
\langle \chi_{m,I}(N)\chi_{m',I}(\bar N)\rangle
\propto \delta(m-m')\, C_1(N,\bar N; m),
\)
so the double integral reduces to a single mass integral:
\begin{equation}
  C_1^{\rm cont}(N,\bar N,\Lambda)
  = \frac{1}{\mu^2}
    \int_{0}^{\infty} \dd m\, g(m)\,
      C_1(N,\bar N; m).
  \label{eq:A_C1_ohm_factorised}
  \end{equation} 
Performing this integral yields
\begin{multline}
C_1^{\rm cont}(N,\bar N,\Lambda)
= -\frac{i\,H\,\Lambda}{2\mu^{3}\mathtt{vol}}\,
e^{-\frac{3}{2}(N+\bar N)}\,
\Bigg[\\
\sinh\!\left(\tfrac{3}{2}(N-N_{0})\right)
\left(-\frac{H\cosh\!\left(\tfrac{3}{2}(N_{0}-\bar N)\right)}
      {\,i H  + (N-2N_{0}+\bar N)\Lambda}+\frac{H\,\sinh\!\left(\tfrac{3}{2}(N_{0}-\bar N)\right)}{\,iH + (N-\bar N)\Lambda\,}\right)\\
+\,H\,\cosh\!\left(\tfrac{3}{2}(N-N_{0})\right)
\left(\frac{\cosh\!\left(\tfrac{3}{2}(N_{0}-\bar N)\right)}{-iH + (N-\bar N)\Lambda}-\frac{i\,\sinh\!\left(\tfrac{3}{2}(N_{0}-\bar N)\right)}{H + i (N-2N_{0}+\bar N)\Lambda}
\right)
\Bigg].
\end{multline}
When passing from these non-local kernels to Markovian ones we let
\(\Lambda\to\infty\) and apply the Sokhotski-Plemelj formula
\begin{equation} \label{eq:Plimeji}
    \frac{\Lambda}{(i \pm \Lambda\alpha)}
  \xrightarrow{\;\Lambda\to\infty\;}
  i\pi\delta(\alpha)\mp\mathcal{PV}\left(\frac{1}{\alpha}\right)\quad\text{with}\quad \int_{0}^{\infty}  \mathcal{PV} \left(\frac{f(x)}{x}\right) \dd x
=\int_{0}^{\infty}  \frac{f(x)-f(0)}{x} \dd x\, .
\end{equation}
After discarding the poles on the zero-measure surface \(N+\bar N=2N_{0}\) we arrive
at the decomposition
\begin{subequations}
\label{eq:A_C_TL}
\begin{align}
  \Re[C_1^{\rm cont}(N,\bar N,\Lambda\to \infty)]
  &=\frac{\pi H^2}{2\mu^3 \mathtt{vol}} e^{-3N}
     \cosh \bigl[3(N-N_{0})\bigr]
     \delta(N-\bar N),
     \label{eq:A_C_TL_re1}\\
  \Im[C_1^{\rm cont}(N,\bar N,\Lambda\to \infty)]
  &=-\frac{
H^{2}\,e^{-3(N+\bar N)}
\left[e^{3\bar N}(N-N_{0}) +e^{3N}(\bar N - N_{0})\right]}{2(N-2N_{0}+\bar N)\,\mu^{3}\mathtt{vol}}\mathcal{PV} \left(\frac{1}{N-\bar N}\right).
     \label{eq:A_C_TL_im1}
\end{align}
\end{subequations}
The first line is an ultralocal white-noise spike; the second is purely imaginary, antisymmetric, and appears as a principal value term. While the imaginary part is finite, the real part grows without bound as the interaction switch-on time is taken to the infinite past
$N_{0}\to -\infty$, due to infinite de Sitter squeezing. To obtain a finite late-time generator we introduce the renormalized coupling
\begin{equation}
\lambda_{1}^{2}\equiv
\lim_{N_{0}\to -\infty}\underline{\lambda}_{1}^{2}e^{-3N_{0}},
\qquad\text{so that}\qquad
\underline{\lambda}_{1}^{2}
=\lambda_{1}^{2}e^{3N_{0}} .
\label{eq:lambdaRenormExp1}
\end{equation}
The real $\delta$ spike dominates the principal value pole which tends to zero with this renormalization scheme. We therefore retain only the symmetric real kernel $\Re[C_1^{\rm cont}(N,\bar N,\Lambda\to \infty)]$ when performing the integral in \eq{eq:A_C_exact}, which yields a completely positive GKLS semigroup with pure dephasing of the form
\begin{equation}
\partial_N \rho
= - i\bigl[\hat K_S(N),\rho\bigr]
  - \frac{\pi \lambda_1^{2}\,\mathtt{vol} \,e^{6 N}}{8 \mu^3} [\hat \phi^3 ,[\hat \phi^3 ,\hat \rho]],
\label{eq:GKLS_app}
\end{equation}
with system Hamiltonian given in \eq{eq:effHamiltonian}. Note the dephasing term picks up an additional factor of $1/2$ since we only integrate $\bar N$ over half of the $\delta(N -\bar N)$ function. In the position basis (field value space) this dissipator acts as
$\partial_{N}\rho(\phi,\phi';N)\vert_{\mathrm{diss}}
= \tfrac{\lambda_1^{2}\,\mathtt{vol}\,e^{6N}}{8\mu^{3}}(\phi^{3}-\phi'^{3})^{2}\rho(\phi,\phi';N)$,
so it leaves the position marginal $\rho(\phi,\phi;N)$ unchanged but
strongly suppresses off-diagonal coherences for $\phi\neq\phi'$.

\subsection*{$\phi^2 \chi^2$ interaction:}
Following \eq{eq:isserlis} and similar steps to the two point derivation we have
\begin{equation}
  C_2^{\rm cont}(N,\bar N,\Lambda)
  = \frac{2}{\mu^2}
    \int_{0}^{\infty} \dd m\, g^2(m)\,
      C_1(N,\bar N; m)^2.
  \label{eq:A_C2_ohm_factorised}
  \end{equation} 
Performing this integral yields
\begin{multline}
C_2^{\rm cont}(N,\bar N,\Lambda)
= \frac{e^{-3 (N + \bar{N})} H \Lambda}{4 \mu^{4} \mathtt{vol
}^2}
\Bigg[
\frac{H \cosh^{2}\!\bigl(\tfrac{3}{2}( \bar{N}-N_{0})\bigr)\, \sinh\bigl(3N - 3N_{0}\bigr)}
     {H - i ( \bar{N}-N_{0}) \Lambda}
\\+ \frac{H \cosh^{2}\!\bigl(\tfrac{3}{2}(\bar{N} - N_{0})\bigr)\, \sinh^{2}\!\bigl(\tfrac{3}{2}(N - N_{0})\bigr)}
       {H - i (N - 2N_{0} + \bar{N}) \Lambda} 
+ \frac{H \sinh^{2}\!\bigl(\tfrac{3}{2}(N - N_{0})\bigr)\, \sinh\bigl( 3(\bar{N}-N_{0})\bigr)}
       {H - i (N - N_{0}) \Lambda} \\
+ \frac{H \sinh\bigl(3N - 3N_{0}\bigr)\, \sinh^{2}\!\bigl(\tfrac{3}{2}(N_{0} - \bar{N})\bigr)}
       {H - i (N_{0} - \bar{N}) \Lambda} 
+ \frac{H \sinh^{2}\!\bigl(\tfrac{3}{2}(N - N_{0})\bigr)\, \sinh^{2}\!\bigl(\tfrac{3}{2}(\bar{N} - N_{0})\bigr)}
       {H - i (N - \bar{N}) \Lambda}
\\- i \cosh^{2}\!\bigl(\tfrac{3}{2}(N - N_{0})\bigr)
  \bigg(
    \frac{H \cosh^{2}\!\bigl(\tfrac{3}{2}(\bar{N} - N_{0})\bigr)}
          {- i H + (N - \bar{N}) \Lambda} 
    - \frac{H \sinh\bigl(3( \bar{N}-N_{0})\bigr)}
           {i H - (N - N_{0}) \Lambda}
    + \frac{H \sinh^{2}\!\bigl(\tfrac{3}{2}(\bar{N} - N_{0})\bigr)}
           {- iH + (N - 2N_{0} + \bar{N}) \Lambda}
  \bigg)
\\+ \sinh\bigl(3N - 3N_{0}\bigr)\, \sinh\bigl(3(\bar{N} - N_{0})\bigr)
\Bigg]. \label{eq:C4_Lambda}
\end{multline}
We can apply the Sokhotski-Plemelj formula \eq{eq:Plimeji} to all terms except the final term \(\propto\Lambda\,
e^{-\frac{3}{2}(N+\bar N)} \sinh\bigl(3N - 3N_{0}\bigr)\, \sinh\bigl(3(\bar{N} - N_{0})\bigr)\) in $C_2^{\rm cont}$ which is infinite in the $\Lambda \to \infty$ limit and analogous to a loop divergence in QFT. This term is removed by redefining the quartic coupling $\underline{\lambda}_2$ with a counter term. The poles of the  Sokhotski-Plemelj formula \eq{eq:Plimeji} are
\begin{equation}
    \alpha_1=N-\bar{N},\quad \alpha_2=N-2 N_0+\bar{N},\quad \alpha_3=N-N_0,\quad\text{and}\quad \alpha_4=\bar{N}-N_0.
\end{equation}
Considering \eq{eq:A_exact_ME} the poles $\alpha_2$ and $\alpha_3$ correspond to equivalent zero-measure integrals $N=\bar{N}=N_0$ and $\alpha_4$ is an initial bath contact term $\bar{N}=N_0$. $\alpha_1$ is the relevant pole for a Markovian master equation and after discarding irrelevant poles we obtain
\begin{equation}
  \Re(C_4^{\rm ohm}(N,\bar N,\Lambda\to \infty))
  =\frac{\pi  H^{2}}{4 \mu^4 \mathtt{vol}^2} e^{-6 N}
     \cosh \bigl[3(N-N_{0})\bigr]
     \delta(N-\bar N),\label{eq:A_C_TL_re2}
 \end{equation}
\begin{multline}
\Im\!\bigl(C_4^{\mathrm{ohm}}(N,\bar N,\Lambda\!\to\!\infty)\bigr)
  = \frac{e^{-3 (N + \bar{N})}  H^{2}}%
  {4 (N - N_{0})(N - 2N_{0} + \bar{N})(\bar{N} - N_{0})\,\mu^{4}\mathtt{vol}^2}\mathcal{PV}\left(\frac{1}{N-\bar{N}}\right)
  \\
  \times
  \Biggl\{
    - (N - N_{0})(N_{0} - \bar{N})^{2}
      \cosh\!\bigl[3(N - N_{0})\bigr]
    + (N - N_{0})^{2}(N_{0} - \bar{N})
      \cosh\!\bigl[3(N_{0} - \bar{N})\bigr]
    \\
    + (N - \bar{N})(N - 2N_{0} + \bar{N})
      \Bigl[
        (N - N_{0}) \sinh\!\bigl[3(N - N_{0})\bigr]
        + (\bar{N} - N_{0}) \sinh\!\bigl[3(N_{0} - \bar{N})\bigr]
      \Bigr]
  \Biggr\}.
\end{multline}
Both of these expressions grow without bound as $N_{0}\to-\infty$ due to exponential de Sitter squeezing. The real part diverges as $e^{-6 N_0}$ and the imaginary part as $e^{-3 N_0}$. To obtain a finite late-time generator we introduce the renormalized coupling 
\begin{equation}
\lambda_2^{2}\equiv\lim_{N_{0}\to-\infty}\underline{\lambda}_2^{2}e^{-6N_{0}},
\qquad\text{so that}\qquad
\underline{\lambda}_2^{2}=\lambda_2^{2} e^{6N_{0}}.
\label{eq:lambdaRenormExp2}
\end{equation}
With this renormalization the imaginary part tends to zero and we derive the master equation
\begin{equation}
\partial_N \rho
= - i\bigl[\hat K_S(N),\rho\bigr]
  - \frac{\pi \lambda_2^{2} e^{6 N}}{64 \mu^4} [\hat \phi^2 ,[ \hat \phi^2 ,\hat \rho]],
\label{eq:GKLS_app2}
\end{equation}
with system Hamiltonian given in \eq{eq:effHamiltonian}. In the position basis the above dissipator is diagonal in $\phi^{2}$, so it
suppresses coherences between configurations with different values of $\phi^{2}$ but leaves intact superpositions related by the $\mathbb{Z}_{2}$ symmetry $\phi\to -\phi$. Consequently, it does not collapse a double-well superposition $\lvert+\phi_{0}\rangle\pm\lvert-\phi_{0}\rangle$
into a single well, since $(+\phi_{0})^{2}=(-\phi_{0})^{2}$.

\subsection*{$\phi \chi^3$ interaction:}
Following \eq{eq:isserlis} and similar steps to the two point derivation we have
\begin{equation}
  C_3^{\rm cont}(N,\bar N,\Lambda)
  = \frac{1}{\mu^2}
    \int_{0}^{\infty} \dd m\, g^3(m)\left[9 C_1(N,\bar N; m)C_1(N, N; m)C_1(\bar N,\bar N; m)
      +6 C_1(N,\bar N; m)^3 \right].
  \label{eq:A_C3_ohm_factorised}
  \end{equation}
The integrated expressions are lengthy and provided in the supplementary Mathematica notebook. We follow a similar procedure to the $C^{\rm cont}_2$ (although loop divergences do not appear in this case) to obtain
 \begin{multline}
 \Re(C_3^{\rm ohm}(N,\bar N,\Lambda\to \infty))=
\frac{e^{-9N} H^{2} \pi}{128\,\mu^{5}}
\Big(131\,\cosh\!\bigl(9(N - N_{0})\bigr)
\\
+\,9\Bigl(
2
- 2\cosh\!\bigl(6(N - N_{0})\bigr)
+ 5\cosh\!\bigl(3(N - N_{0})\bigr)
\Bigr)
\Big).
\end{multline}
The expression for the imaginary part is given in the supplementary Mathematica notebook and grows as $e^{-6 N_0}$ as $N_0\to -\infty$ whilst the real part grows as $e^{-9 N_0}$. To obtain a finite result, we renormalise the coupling $\underline{\lambda}_3$ as 
\begin{equation}
\lambda^{2}\equiv\lim_{N_{0}\to-\infty}\underline{\lambda}_3^{2}e^{-9N_{0}},
\qquad\text{so that}\qquad
\underline{\lambda}_3^{2}=\lambda^{2} e^{9N_{0}}.
\label{eq:lambdaRenormExp3}
\end{equation}
We arrive at the GKLS equation
\begin{equation}
\partial_N \rho
= - i\bigl[\hat K_S(N),\rho\bigr]
  - \frac{131 \pi \lambda^{2} e^{6 N}}{512 \mu^5 \mathtt{vol}} [\hat \phi,[ \hat \phi,\hat \rho]],
\label{eq:GKLS_app3}
\end{equation}
with system Hamiltonian given in \eq{eq:effHamiltonian}. This corresponds to standard position dephasing, with
$\partial_{N}\rho(\phi,\phi';N)\vert_{\mathrm{diss}}
= -\tfrac{131 \pi \lambda^{2} e^{6N}}{512 \mu^{5} \mathtt{vol}}(\phi-\phi')^{2}\rho(\phi,\phi';N)$,
which damps all off-diagonal coherences in the $\phi$ basis while leaving
the position marginal $\rho(\phi,\phi;N)$ unchanged. Upon identifying $\lambda$ with the renormalised coupling used in the main text, this expression reproduces Eq.~\eqref{eq:GKLS_phi_only} with the rate $\Gamma_\phi(N)$ defined there.

\section{Late-time Gaussian SSE estimate of bath-induced barrier hopping rate}
\label{app:GaussSSEArrhenius}

In this appendix we estimate a bath-induced Arrhenius-type mean first passage time from the false vacuum $F$ to the true vacuum $T$ of the form \cite{ref:chandlerbook}
\begin{equation}
    \left\langle N_{F \rightarrow T}\right\rangle \simeq \frac{2\pi}{\sqrt{V''_{\rm eff}(\phi_F)\,\big|V''_{\rm eff}(\phi_{\rm top})\big|}}\,
    \exp\!\left(\frac{\Delta V^{\rm eff}_{FT}}{D_{\rm bath}}\right),
\label{eq:appRateDef}
\end{equation}
where $V_{\rm eff}$ is the overdamped effective potential, $\Delta V^{\rm eff}_{FT}=V_{\rm eff}(\phi_{\rm top})-V_{\rm eff}(\phi_F)$, and $D_{\rm bath}$ is the diffusion strength in $N$-time.

We approximate late-time ($N\gg 1$) stochastic Schr\"odinger trajectories by a Gaussian Wigner state
\begin{equation}
W(z,N)=\frac{1}{2\pi\sqrt{\det\Sigma(N)}}\,
\exp\!\left[-\frac{1}{2}\big(z-Z(N)\big)^{\!\top}\Sigma^{-1}(N)\big(z-Z(N)\big)\right],
\label{eq:appGaussAnsatz}
\end{equation}
with $z=(\phi,\pi_{\phi})^{\top}$, center $Z=(\ev{\phi},\ev{\pi_{\phi}})^{\top}$, and $2\times 2$ covariance matrix $\Sigma$.
The Weyl-form SSE implies the Gaussian parameter equations \cite{MyPaper}
\begin{equation}
\dd Z=\Omega\nabla H\, \dd N+\sum_{j}\Big[\Omega\,\mathrm{Im}(L_j\nabla\bar L_j)\,\dd N+\Big(2 \Sigma\nabla L_j^{R}-\Omega\nabla L^{I}_j\Big)\,\dd W_j\Big],
\label{eq:appSSEZ}
\end{equation}
\begin{equation}
\frac{\dd \Sigma}{\dd N}=\Omega H''\Sigma-\Sigma H''\Omega-\sum_{j}\Big[4\,\Sigma\,\mathrm{Re}\!\big(\nabla\bar L_j\,\nabla L^{\top}_j\big)\Sigma+\Omega\,\mathrm{Re}\!\big(\nabla\bar L_j\,\nabla L^{\!\top}_j\big)\Omega\Big],
\label{eq:appSSESigma}
\end{equation}
where $\Omega=\bigl(\begin{smallmatrix}0&1\\-1&0\end{smallmatrix}\bigr)$. Near a minimum we take $V(\phi)\simeq \tfrac12m_{F}^2\phi^2$. The Hamiltonian is
\begin{equation}
H(N)\simeq \frac{e^{-3N}}{2H\,\mathtt{vol}}\pi_{\phi}^2+\frac{e^{3N}\mathtt{vol}}{H}\,\frac12m_{F}^2\phi^2,
\label{eq:appHam}
\end{equation}
so the deterministic drift of the phase-space center is
\begin{equation}
\mu_\phi(N,Z)=\frac{e^{-3N}}{H\,\mathtt{vol}}\,\pi_{\phi},\qquad
\mu_{\pi}(N,Z)=-\frac{e^{3N}\mathtt{vol}}{H}\,m_{F}^2\phi .
\label{eq:appDrift2D}
\end{equation}
Keeping only the bath channel, we take a single Lindblad operator linear in $\phi$,
\begin{equation}
    L_{\rm bath}(N)=\sqrt{\Gamma_{\rm bath}(N)}\,\phi,
    \qquad
    \Gamma_{\rm bath}(N)=\frac{131\pi\,\lambda^2}{256\,\mu^5\,\mathtt{vol}}\,e^{6N}.
\label{eq:appLbath}
\end{equation}

Solving Eq.~\eqref{eq:appSSESigma} for the instantaneous fixed point ($\dd \Sigma/\dd N=0$) in the harmonic approximation and taking the late-time limit yields an It\^o noise vector for the phase-space center with
\begin{equation}
A_\phi^{(\infty)}=0,
\qquad
A_{\pi}^{(\infty)}=\frac{\sqrt{131\pi}}{16}\,
\frac{\lambda e^{3N}}{\sqrt{\mathtt{vol}\,\mu^5}}\,\dd W .
\label{eq:appNoiseBathOnly}
\end{equation}
The bath therefore produces asymptotically momentum diffusion with the characteristic $e^{3N}$ enhancement.

The coupled Langevin system for the center is
\begin{equation}
\dd\phi=\frac{e^{-3N}}{H\,\mathtt{vol}}\,\pi_{\phi}\,\dd N,\qquad
\dd \pi_{\phi}=-\frac{e^{3N}\mathtt{vol}}{H}\,m_{F}^2\phi\,
\dd N+\frac{\sqrt{131\pi}}{16}\,
\frac{\lambda e^{3N}}{\sqrt{\mathtt{vol}\,\mu^5}}\,\dd W .
\label{eq:appLangevin2D_bath}
\end{equation}
Combining these gives the second-order stochastic equation
\begin{align}
\phi''+3\phi'+\frac{m_{F}^2}{H^2}\phi&=\frac{e^{-3N}}{H\,\mathtt{vol}}\;\frac{\dd}{\dd N}\left(\int^N A_{\pi}^{(\infty)}\right) \nonumber\\
&=\sqrt{\frac{\lambda^2 131\pi}{256 H^2\mathtt{vol^3}\,\mu^5}}\;\xi(N),
\label{eq:appSecondOrder_bath}
\end{align}
where $\xi(N)\equiv \dd W/\dd N$ is white noise.

If the field has a low mass compared to the Hubble mass, at late times the dynamics is overdamped, so we drop $\phi''$ and obtain an effective one-dimensional bath-driven equation
\begin{equation}
\dd\phi\simeq -\frac{m^2_F}{3H^2}\,\dd N+\sqrt{2D_{\rm bath}}\,\dd W,
\qquad
D_{\rm bath}=\frac{131\pi}{4608}\,
\frac{\lambda^2}{H^{2}\,\mathtt{vol}^{3}\,\mu^{5}}.
\label{eq:appSmolBath}
\end{equation}

The $e^{3N}$ growth in $A_\pi^{(\infty)}$ precisely cancels the $e^{-3N}$ kinematic suppression in the $\phi$ equation, leaving a constant late-time diffusion strength for $\phi$.

The overdamped effective potential is
\begin{equation}
V_{\rm eff}(\phi)=\frac{V(\phi)}{3H^2},
\qquad
V_{\rm eff}''(\phi)=\frac{V''(\phi)}{3H^2},
\qquad
\Delta V^{\rm eff}_{FT}=\frac{V(\phi_{\rm top})-V(\phi_F)}{3H^2}.
\label{eq:appVeffDefs}
\end{equation}
The Eyring--Kramers estimate Eq~\eqref{eq:appRateDef} gives
\begin{equation}
\left\langle N_{F \rightarrow T}\right\rangle \simeq \frac{6 \pi H^2}{\mu^2} \sqrt{\frac{\beta_4+\beta_3}{2 \beta_4}} \exp \left[\frac{128}{131 \pi} \frac{3 \beta_4+\beta_3}{\left(\beta_4+\beta_3\right)^3} \frac{\mathtt{vol}^3 \mu^9}{\lambda^2}\right]
\end{equation}

\section{Adiabatic theorem for an oscillator in de Sitter} \label{app:adiabatic}

Consider the time-dependent quadratic Hamiltonian
\begin{equation}
K_S(t) = \frac{\pi_\phi^2}{2\,e^{3Ht}} + \frac{e^{3Ht}}{2}\,m^2 \phi^2,
\label{eq:KS_definition}
\end{equation}
with $N \equiv Ht$. This can be written in the general form of a parametric oscillator,
\begin{equation}
\hat K_S(t) = \frac{\hat\pi^2}{2M(t)} + \frac{M(t)}{2}\,\omega^2 \hat\phi^2,
\qquad
\text{where } M(t) = e^{3Ht}, \;\; \omega = m.
\label{eq:parametric_oscillator}
\end{equation}
The time derivative in the Schrödinger picture is
\begin{equation}
\dot{\hat K}_S(t)
= -\frac{\dot{M}}{2 M^2} \hat\pi^2 + \frac{\dot{M}}{2} \omega^2\hat \phi^2
= \frac{\dot{M}}{2 M} \omega \left(\hat a^2 + \hat a^{\dagger 2}\right),
\label{eq:Kdot}
\end{equation}
where $\hat a$ and $ \hat a^\dagger$ are the instantaneous ladder operators associated with $M(t)$ and $\omega$. A sufficient condition for adiabatic evolution is given in \cite{comparat2009general} as
\begin{equation}
\frac{\left| \langle m(t) | \dot{\hat K}_S(t) | n(t) \rangle \right|}{\left[E_n(t) - E_m(t)\right]^2} \ll 1 
\qquad \text{for all } m \neq n,
\label{eq:adiabatic_condition}
\end{equation}
here $n$ and $m$ label the eigenstates with energies $E_n$ and $E_m$ respectively. The nonvanishing matrix elements of $\dot{\hat K}_S$ occur only between states differing by two quanta. Evaluating these gives the adiabatic parameters
\begin{equation}
\frac{\left|\langle n+2| \dot{\hat K}_S | n\rangle\right|}{\left(E_n - E_{n+2}\right)^2}
= \frac{3 H}{8 m} \sqrt{(n+1)(n+2)}, 
\qquad
\frac{\left|\langle n-2| \dot{ \hat K}_S | n\rangle\right|}{\left(E_n - E_{n-2}\right)^2}
= \frac{3 H}{8 m} \sqrt{n(n-1)},
\label{eq:adiabatic_parameters}
\end{equation}
and zero for all other \( m \neq n \).
In particular, the ground-state (vacuum) mixing parameter is
\begin{equation}
\frac{\left|\langle 2| \dot{ \hat K}_S | 0\rangle\right|}{\left(E_0 - E_2\right)^2}
= \frac{3 \sqrt{2}}{8} \frac{H}{m}.
\label{eq:groundstate_mixing}
\end{equation}
Hence, the evolution is adiabatic provided that
\begin{equation}
\frac{m}{H} \gg 1\, .
\label{eq:adiabatic_condition_final}
\end{equation}

\section{Additional simulations}

\FloatBarrier

To keep the main text concise, we omitted several simulation results that are not essential for understanding the effects of non-adiabatic dynamics, decoherence, and false-vacuum populations. In this appendix, we present a selection of these additional simulations.

\subsection{$\phi^2 \chi^2$ four point interactions}

\begin{figure}
    \centering
    \begin{subfigure}[t]{0.32\textwidth}\centering
        \panel{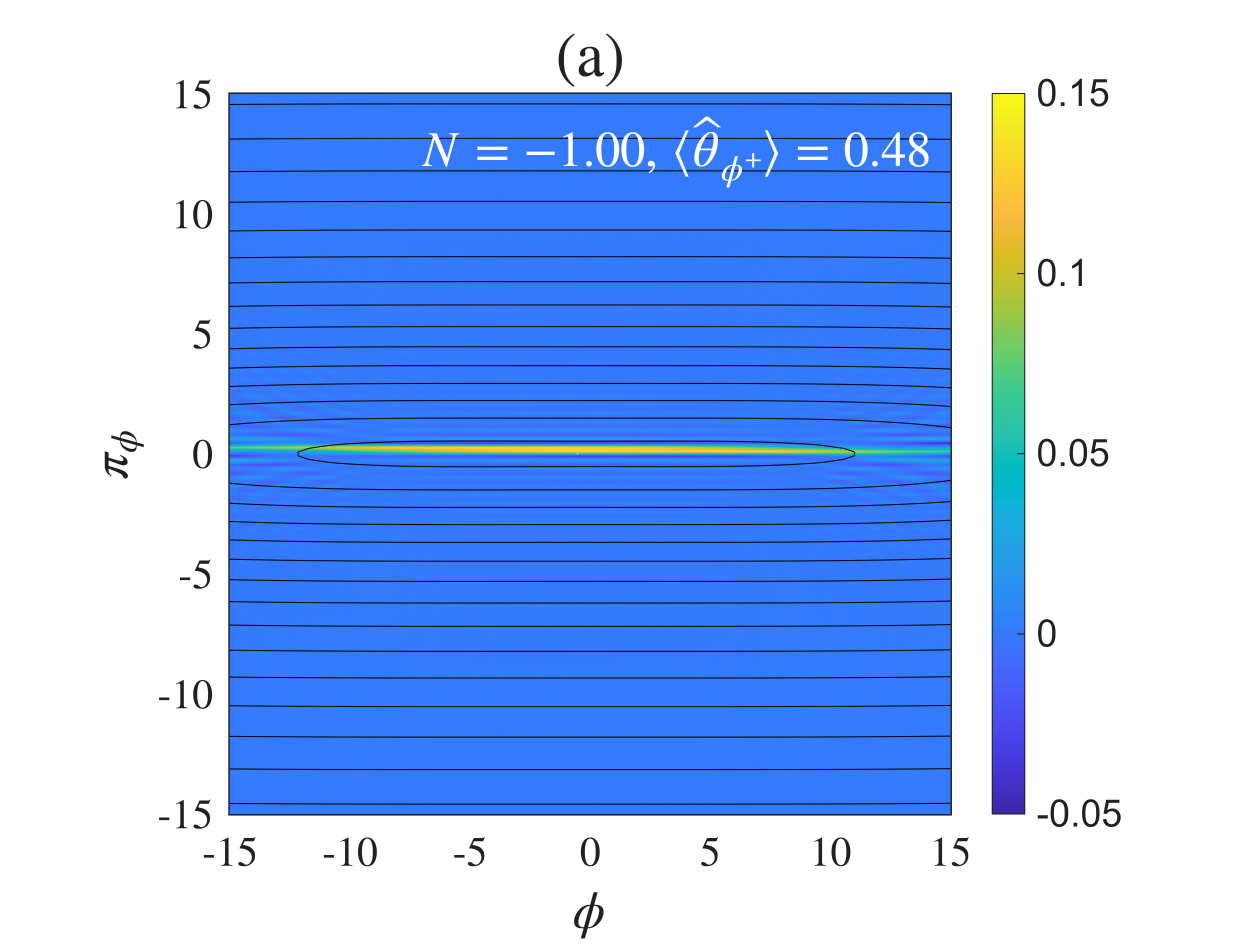}
    \end{subfigure}\hfill
    \begin{subfigure}[t]{0.32\textwidth}\centering
        \panel{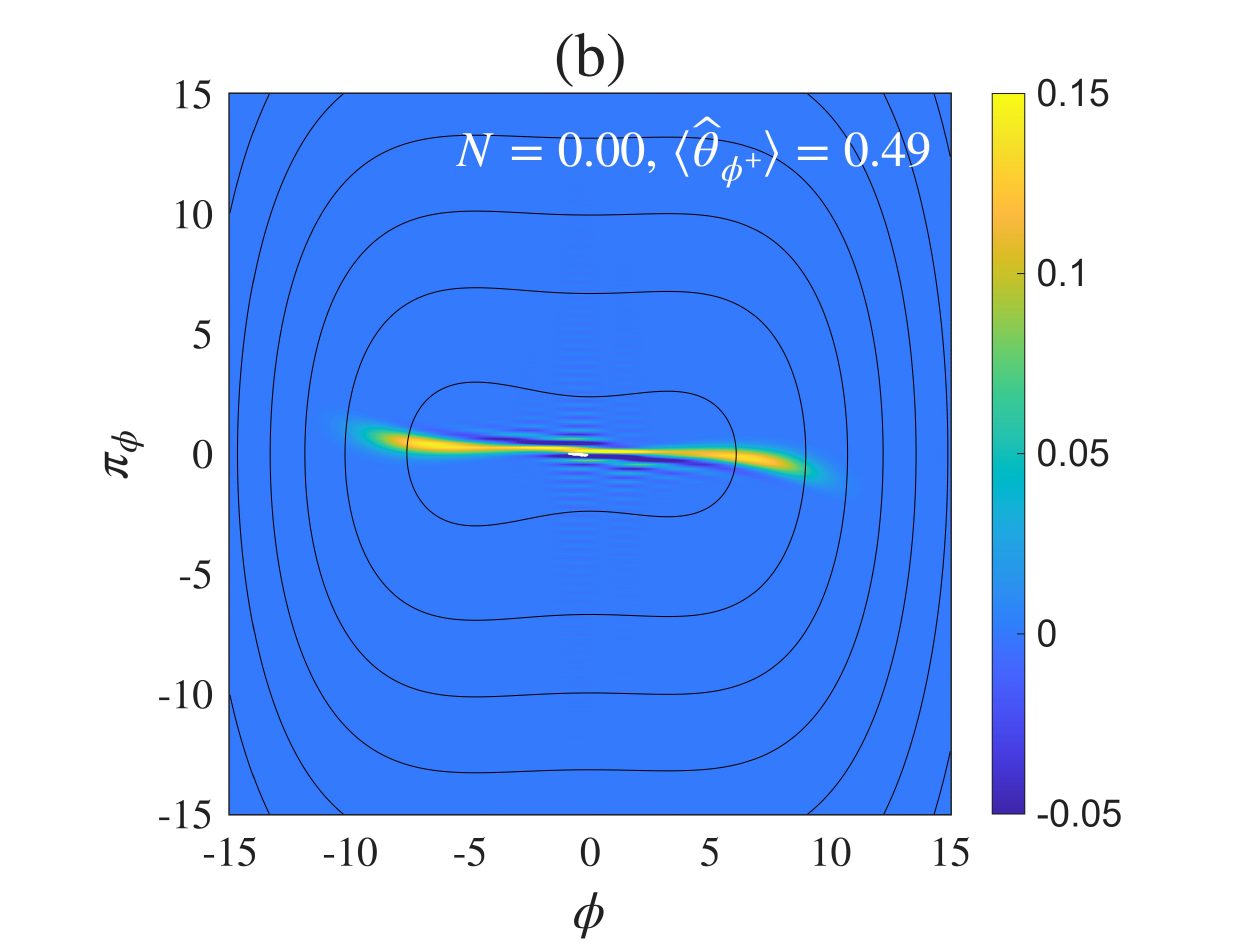}
    \end{subfigure}\hfill
    \begin{subfigure}[t]{0.32\textwidth}\centering
        \panel{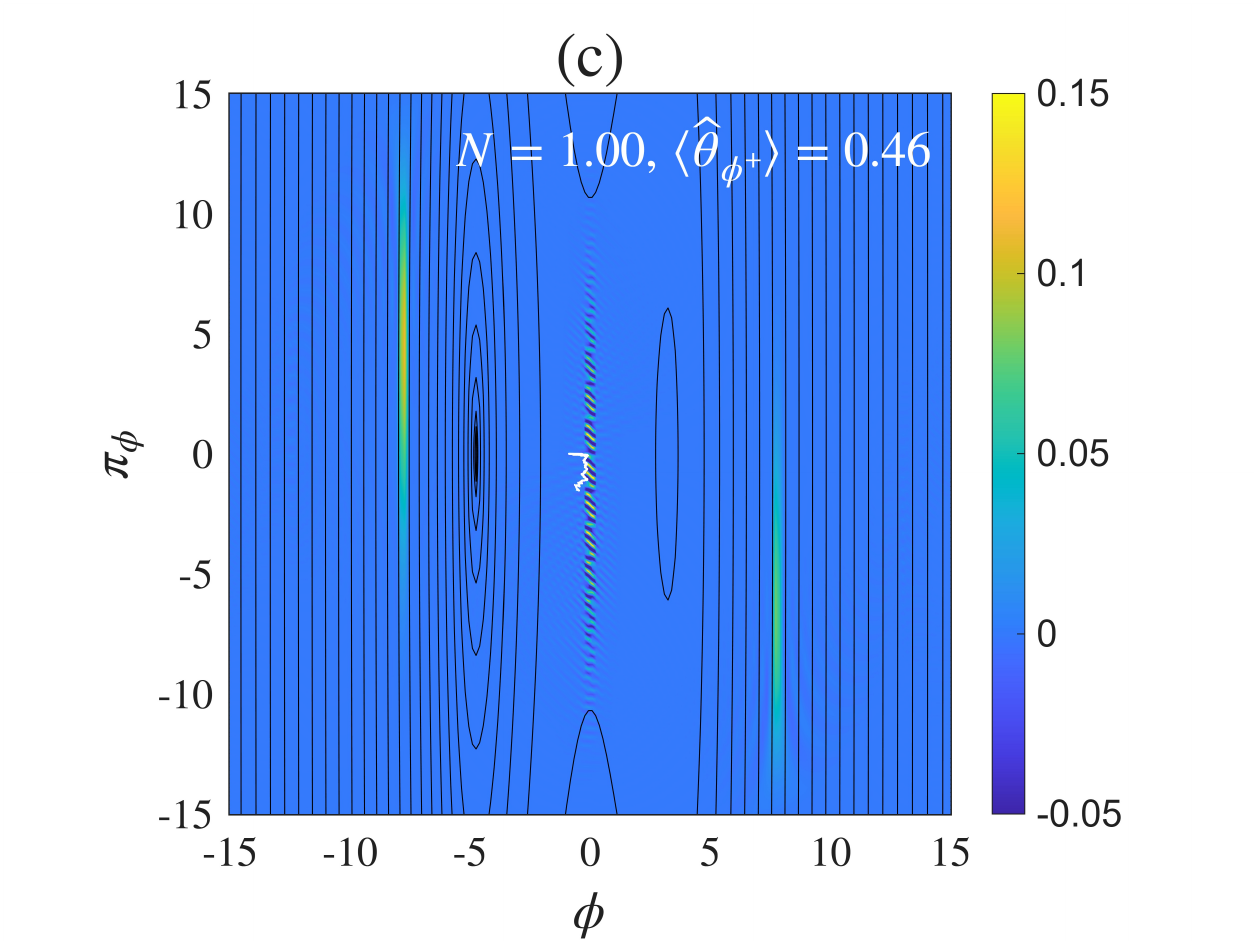}
    \end{subfigure}
    \caption{\small
    Wigner functions for the $\mathbb{Z}_2$ symmetric SSE evolution Eq.~\eqref{eq:SSE_Z2}. The columns correspond to different snapshot times. Black curves are equal-energy contours of $K_S(N)$. Axes are $\phi$ and $\pi_\phi$.
    In-panel text reports the $e$-fold $N$ and the false vacuum projector expectation value. The sonified video corresponding to these plots is available
    \href{https://youtube.com/shorts/tQEgCfBLNT0}{via this link}. The white line corresponds to the prior evolution of the phase space expectation values $\ev*{\hat \phi}$ and $\ev*{\hat \pi_{\phi}}$. }
    \label{fig:WignerSSEX2}
\end{figure}

We also investigated a $\mathbb{Z}_2$–symmetric stochastic unraveling in which the environment couples quadratically to the spectator field, leading to the nonlinear SSE
\begin{equation}
\dd \ket{\psi}
= - i\hat K_S \ket{\psi}\,\dd N
- \frac{\pi \lambda^{2} e^{6N}}{64 \mu^4}
\Big( \hat \phi^2 - \langle \hat \phi^2 \rangle \Big)^{2} \ket{\psi}\,\dd N
+ \frac{\,\lambda\, e^{3N}}{4 \mu^2}\sqrt{\frac{\pi}{2}}
\Big( \hat \phi^2 - \langle \hat \phi^2 \rangle \Big)\ket{\psi}\,\dd W ,
\label{eq:SSE_Z2}
\end{equation}
with Lindblad operator proportional to $\hat\phi^2$. In contrast to the $\hat L\propto\hat\phi$ channel discussed in the main text, this $\mathbb{Z}_2$–symmetric dynamics dephases in the $\hat\phi^2$ pointer basis and therefore preserves the $\phi\to -\phi$ symmetry: superpositions of states with opposite sign but equal magnitude of $\phi$ are not distinguished by the bath. As a result, the $\phi^2$–coupled environment efficiently suppresses coherences between configurations with different $|\phi|$, but it does not by itself drive vacuum selection between the true and false minima in an asymmetric double well. This behaviour is clearly visible in the Wigner snapshots of Fig.~\ref{fig:WignerSSEX2}, where the state becomes strongly mixed and increasingly classical in $|\phi|$ while retaining approximate symmetry between positive and negative field values, in marked contrast to the $\hat L\propto\hat\phi$ trajectories that rapidly localize into a single well.

\subsection{Non-Markovian single spectator environment}
\label{sec:nonmarkovian_single_spectator}
\begin{figure}

    \begin{subfigure}[t]{0.32\textwidth}\centering
        \panel{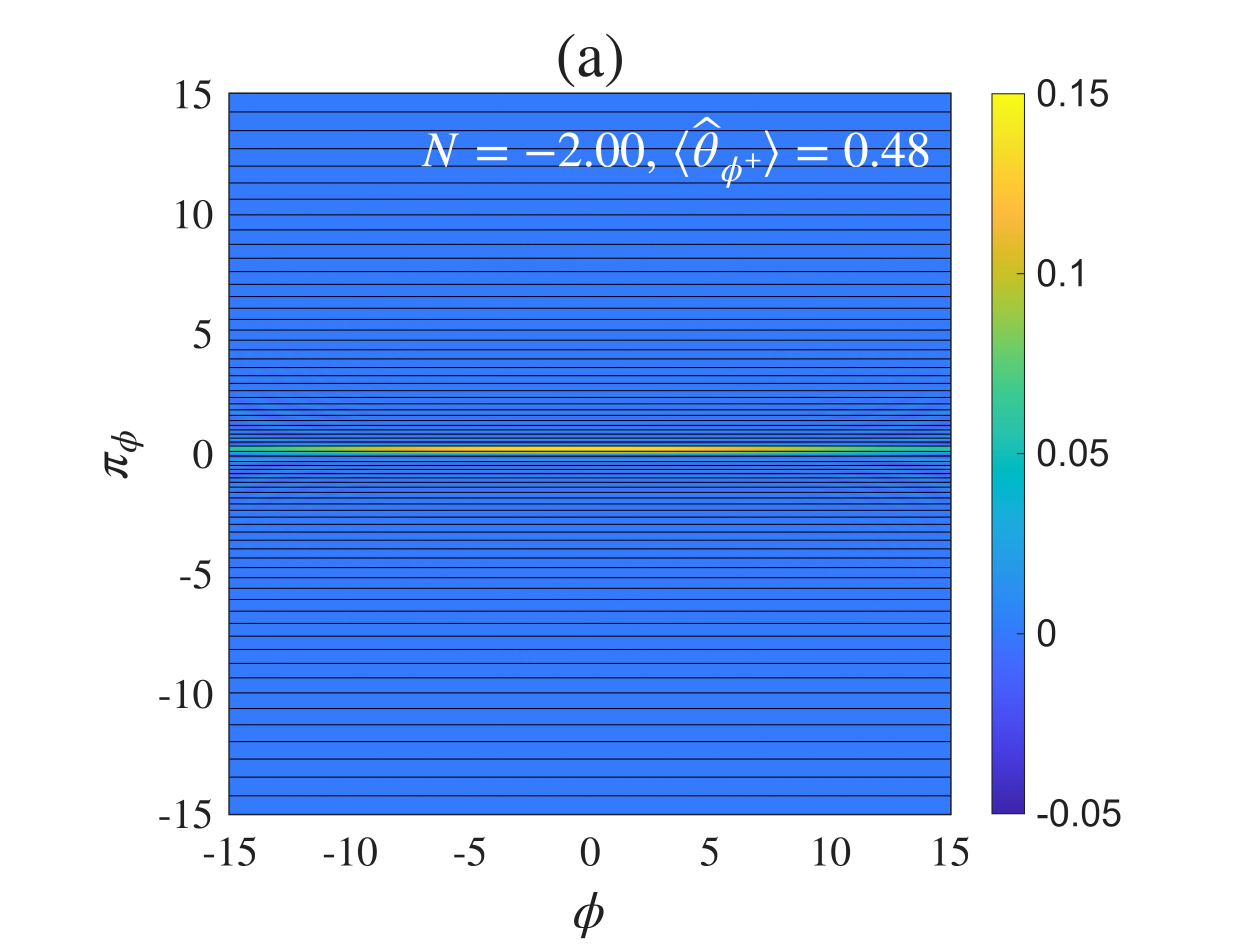}
    \end{subfigure}\hfill
    \begin{subfigure}[t]{0.32\textwidth}\centering
        \panel{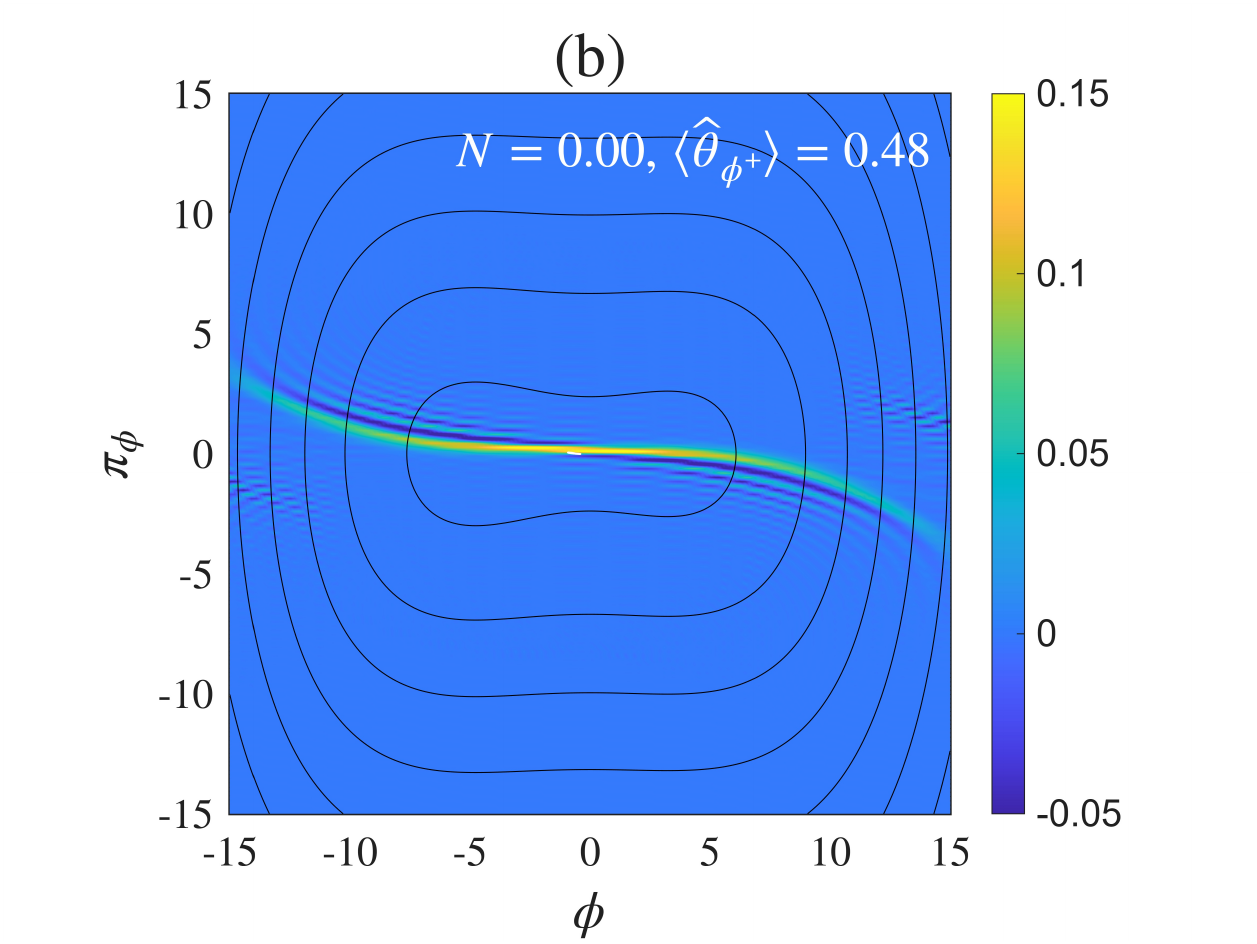}
    \end{subfigure}\hfill
    \begin{subfigure}[t]{0.32\textwidth}\centering
        \panel{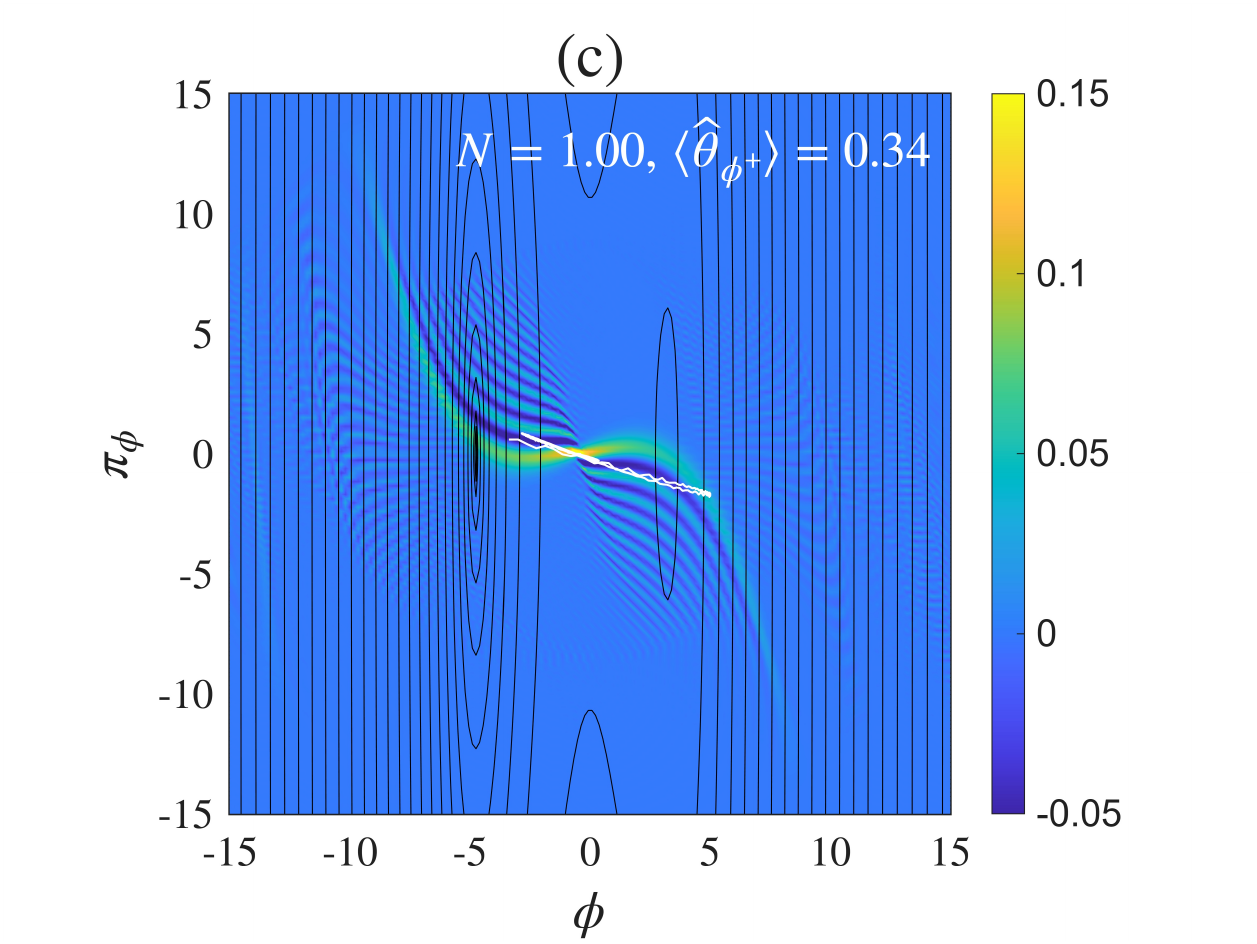}
    \end{subfigure}
    \caption{\small
    Wigner functions for the NMQSD (Eq.~\eqref{eq:NMQSD}) trajectory with coupling $\lambda_3=100$ and spectator frequency $\omega_{\chi}=10$.
    Columns correspond to different snapshot times. Black curves are equal-energy contours of $K_S(N)$.
    Axes are $\phi$ and $\pi_\phi$.
    In-panel text reports the $e$-fold $N$ and the false vacuum projector expectation value. The sonified videos corresponding to these plots are available via these links: \href{https://youtube.com/shorts/9x7ke1iWXnQ}{this link}. The white line corresponds to the prior evolution of the phase space expectation values $\ev*{\hat \phi}$ and $\ev*{\hat \pi_{\phi}}$. }
    \label{fig:WignerNMQSD}
\end{figure}
A convenient way to realise a genuinely non-Markovian dynamics is to couple the spectator field to a \emph{single} massive environment field \(\chi\) via the cubic interaction \(\phi\chi^{3}\), and then apply the non-Markovian quantum state diffusion (NMQSD) formalism of Diósi, Gisin and Strunz~\cite{diosi1998non,diosi2014general}. In this case the reduced dynamics of the system field is unraveled by the nonlinear, norm-preserving NMQSD equation,
\begin{equation}
\frac{\partial}{\partial N}\,\ket{\psi(z^*)}
= \left[
   -\,i \hat K_{ S}(N)
   +\frac{\lambda e^{3 N}}{H}\bigl(\hat \phi- \ev*{\hat \phi} \bigr)\,z_N^*
   -\frac{\lambda e^{3 N}}{H}\bigl(\hat \phi - \ev*{\hat \phi}\bigr)
    \!\int_{N_0}^N\!\mathrm{d}\bar N\;
      C(N,\bar N)\,
      \frac{\delta}{\delta z_{\bar N}^*}
  \right]\,\ket{\psi(z^*)}, \label{eq:NMQSD}
\end{equation}
where \(z_N \equiv z(N)\) is a complex Gaussian process with zero mean and coloured correlations
\begin{equation}
  \mathbb{E}\!\big[z_N\big]=0,
  \qquad
  \mathbb{E}\!\big[z_N z_{\bar N}\big]=0,
  \qquad
  \mathbb{E}\!\big[z_N z_{\bar N}^*\big]=C(N,\bar{N}).
\end{equation}
The same kernel \(C(N,\bar N)\) therefore governs both the non-local memory term and the noise correlations.

The non-local memory kernel \(C(N,\bar N)\) is fixed by the spectator-field correlators. Writing the total interaction as
\(\hat H_{\rm int}(N) = \tilde \lambda_3\,\hat\phi \,\hat\chi_I^{3}(N)\), one finds
\begin{equation}
  C(N,\bar N)
  \;=\;
  \tilde{\lambda}_3^{2}\,
  \left[
    \big\langle 0_{N_0}\big|\hat\chi_I^{3}(N)\,\hat\chi_I^{3}(\bar N)\big|0_{N_0}\big\rangle
    - \big\langle 0_{N_0}\big|\hat\chi_I^{3}(N)\big|0_{N_0}\big\rangle
      \big\langle 0_{N_0}\big|\hat\chi_I^{3}(\bar N)\big|0_{N_0}\big\rangle
  \right].
\label{eq:C_NNbardeSitter}
\end{equation}
Applying the renormalization scheme in Eq.~\eqref{eq:lambdaRenormExp3} yields a stationary correlator that depends only on the e-fold difference \(\Delta N = N-\bar N\),
\begin{equation}
C(N,\bar N)=
\frac{3\lambda_3^{2}}{128\,H^{3}\,\omega_\chi^{3}}
\Bigl[
9\cos\bigl(\omega_\chi \Delta N\bigr)
+\cos\bigl(3\omega_\chi \Delta N\bigr)
\Bigr],
  \qquad \omega_{\chi}\equiv\frac{m_{\chi}}{H},
\label{eq:C_cosine_env}
\end{equation}
so that \eqref{eq:C_NNbardeSitter} provides an explicit non-local kernel \(C(N,\bar N)\) for use in the NMQSD equation and the corresponding non-Markovian master equation. In the Wigner snapshots of Fig.~\ref{fig:WignerNMQSD} we can see that the periodic non-Markovian driving can lead to periodic transitions between vacua.

\end{document}